\documentclass[aps, reprint, superscriptaddress, showpacs, showkeys]{revtex4-1}

\usepackage[T1]{fontenc}
\usepackage{hyperref}

\usepackage{dcolumn}% Align table columns on decimal point
\usepackage{tabularx}
\usepackage{amsfonts}
\usepackage{chemist}
\usepackage{amssymb,amsmath, bm}
\usepackage{subfigure}
\usepackage{epsfig}
\usepackage{color}
\newcommand{\un}{\: \mathrm}

\usepackage{xspace}
\newcommand{\Edot}{\dot{E}_0\xspace}
\newcommand{\HSm}{HS-m\xspace}
\newcommand{\SM}{SM\xspace}
\newcommand{\TM}{TM\xspace}
\newcommand{\HSp}{HS-p\xspace}

\begin{document}

\title{Role of particle aggregation on the structure of dried colloidal silica layers}

\author{Arnaud Lesaine}
\affiliation{Laboratoire FAST, Univ. Paris-Sud, CNRS, Universit\'e Paris-Saclay, F-91405, Orsay, France.}
\affiliation{SPEC, CEA, CNRS, Universit\'e Paris-Saclay, 91191 Gif-sur-Yvette, France.}
\author{Daniel Bonamy}
\affiliation{SPEC, CEA, CNRS, Universit\'e Paris-Saclay, 91191 Gif-sur-Yvette, France.}
\author{Cindy L. Rountree}
\affiliation{SPEC, CEA, CNRS, Universit\'e Paris-Saclay, 91191 Gif-sur-Yvette, France.}
\author{Georges Gauthier}
\affiliation{Laboratoire FAST, Univ. Paris-Sud, CNRS, Universit\'e Paris-Saclay, F-91405, Orsay, France.}
\author{Marianne Impéror-Clerc}
\affiliation{Université Paris-Saclay, CNRS, Laboratoire de Physique des Solides, 91405, Orsay, France.}
\author{Véronique Lazarus}
\affiliation{Laboratoire FAST, Univ. Paris-Sud, CNRS, Universit\'e Paris-Saclay, F-91405, Orsay, France.}
\affiliation{IMSIA, ENSTA ParisTech, CNRS, CEA, EDF, Universit\'e Paris-Saclay, 828 bd des Mar\'echaux, 91762 Palaiseau Cedex, France.}

\begin{abstract}
The process of colloidal drying gives way to particle self-assembly in numerous fields including photonics or biotechnology. Yet, the mechanisms and conditions driving the final particle arrangement in dry colloidal layers remain elusive. Here, we examine how the drying rate selects the nanostructure of  thick dried layers in four different suspensions of silica nanospheres. Depending on particle size and dispersity, either an amorphous arrangement, a crystalline arrangement, or a rate-dependent amorphous-to-crystalline transition occurs at the drying surface. Amorphous arrangements are observed in the two most polydisperse suspensions while crystallinity occurs when dispersity is lower. Counter-intuitively in the latter case, a higher drying rate favors ordering of the particles. To complement these measurements and to take stock of the bulk properties of the layer, tests on the layer porosity were undertaken. For all suspensions studied herein, faster drying yields denser dry layers. Crystalline surface arrangement implies large bulk volume fraction ($\sim 0.65$) whereas amorphous arrangements can be observed in layers with either low (down to $\sim 0.53$) or high ($\sim 0.65$) volume fraction. Lastly, we demonstrate via  targeted additional experiments and SAXS measurements, that the packing structure of the layers is mainly  driven by the formation of aggregates and their subsequent packing, and not by the competition between Brownian diffusion and convection.  This highlights that a second dimensionless ratio in addition to the Peclet number should be taken into account, namely the aggregation over evaporation timescale. 
\end{abstract}

%{Colloids}
%{Self-assembly in material synthesis & processing}

\maketitle

\section{Introduction}

As a colloidal suspension dries, the evaporation of the liquid phase leads to the formation of a solid layer. Such dried colloidal layers are found in biological and geological systems, such as bloodstains\cite{AtMoDo13} and clay.\cite{TaCuTa11} They are also involved in many industrial processes: printing and painting,\cite{Rou13} manufacturing protective or decorative coatings,\cite{KedRou10} preparing ceramics and glasses with tunable properties,\cite{BriSch90} fabricating photonics crystals and artificial opals\cite{Col01,VlaBoStu01} etc. In this context, understanding, predicting and controlling the mechanisms driving particle self-assembly represents a major challenge. Obtaining defect-free colloidal crystals -- which can be done via drying\cite{BlaChoGra00} -- is a key milestone for \textit{e.g.}\ photonic band gap applications. Moreover, dry colloidal layers can contain unwanted cracks,\cite{GoeNakDut15} the presence of which should be limited. The structure of the layer selects the mechanical properties,\cite{LesBonGau18,Les18} which in turn drastically affect the cracking pattern.\cite{GauLazPau10,LazPau11,GioPau14,Laz17} In this respect, the drying rate is an important control parameter.\cite{CadHul02,GauLazPau10,BouGioPau14}

Various experiments have thus been developed to understand how colloidal particles self-assemble during drying. Large scale crystalline arrangements, as well as an order-to-disorder transition, were first observed\cite{NarWanLin04,BigLinNgu06,MarGelLoh11,MonLeq11,AskSefKou13} on thin layers (with thickness ranging from one to few particle diameters) obtained by evaporating suspension droplets. Similar observations were also reported\cite{GoeCleRou10,InOsKa16,NoiMarDev17,YouWanWan17} in unidirectional drying experiments in Hele-Shaw cells of small thickness (from $\sim$ 1\cite{YouWanWan17, NoiMarDev17} to $\sim 100$\cite{GoeCleRou10,InOsKa16} bead diameters). In these experiments, the velocity at which particles migrate toward the contact line (in sessile drop geometry) or liquid/solid interface (in Hele-Shaw geometry) arose as the main parameter controlling the transition between order and disorder. Still, the underlying mechanisms remain elusive;  in some cases, higher velocity yields better ordering,\cite{AskSefKou13,NoiMarDev17,PirLazGau16} while in some others it prevents ordering.\cite{MarGelLoh11,YouWanWan17}

Previously, Piroird et al.\cite{PirLazGau16} analyzed the structure of \textit{millimeter-thick} layers obtained by unidirectional drying of suspensions of monodisperse silica nanospheres. Layers dried at high evaporation rates displayed a crystalline structure on the drying surface and a denser packing of the silica particles in their bulk. On the other hand, slowly-dried layers had an amorphous surface and lower volume fraction. These observations were interpreted by conjecturing aggregate formation: When layers dry slowly, they remain long enough in a concentrated, liquid state for silica particles to aggregate. These aggregates then preclude the crystallization and tighter packing observed in fast-dried layers. 

Here, we extend this previous study by exploring the role of particle size and dispersity on the final structure of the dried layers. Experiments were conducted on four different silica suspensions.  Before conducting the drying experiments, small angle X-ray
scattering (SAXS) measurements enabled the characterization of the particle
diameter and its standard deviation for each silica
suspension. Section 2 details the experimental toolboxes used to achieve these experiments. Depending on the system, crystalline arrangements, amorphous arrangements, or a rate-dependent amorphous-to-crystalline transition are observed at the surface of the dry layers (Sec.\ 3); both faster drying and lower polydispersity promote higher ordering. As in,\cite{PirLazGau16} these changes in surface ordering go along with changes in bulk volume fraction: crystalline surface arrangement is associated with high bulk volume fraction, while amorphous surface structures are observed in either dense or loose layers. SAXS measurements on a suspension subjected to a holding period experiment confirm that, as conjectured in,\cite{PirLazGau16} aggregate formation is the key mechanism responsible for the observed disorder-to-order transition (Sec.\ 4). These results are finally discussed in Sec.\ 5.  

\section{Methods}

\subsection{Specifications and SAXS characterization of the colloidal suspensions}

Ludox suspensions consist of a mass fraction $\phi_m$ of small spheres of amorphous silica ($a- SiO_2$) in water. These suspensions are stabilized using sodium as a counter-ion to prevent aggregationin the store bought concentration. Herein, the Ludox suspensions are used directly as supplied by Sigma-Aldrich (\textit{i.e.} that is, no treatment such as washing or filtration occurs before the drying experiments). The study herein concerns four different batches from three different types of Ludox: Ludox SM, Ludox HS-40 (two batches of Ludox HS-40 were used), and Ludox TM were used. These Ludox types are simply referred to as SM, HS and TM in the following. Table\ \ref{tab:ludox_spec} provides a summary of the manufacturer's  (Grace Davidson) specifications. They allow a rough estimate of the colloidal particle sizes.

%\begin{table}[htbp]
\begin{table}[h]
    \caption{Four first rows: specifications supplied by the producer of the three Ludox types used in this study (silica mass fraction $\phi_m^{sil}$, specific area of the particles S, sodium mass fraction $\phi_m^{Na}$, and pH). Fifth row: particle radius, $R$, inferred from specific area, $S$, using $R = 3/(\rho_{sil}S)$ and $\rho_{sil} = 2.26 \un{g/cm^3}$ (see Sec.\ 2.3)}
  \centering
    \begin{tabular}{cccc}
    \hline
		  & SM & HS & TM \\
		\hline 
	Silica ($\phi_m^{sil}$) & $0.29 - 0.31$ & $0.39 - 0.41$ & $0.39 - 0.41$ \\
	$S$ ($\mathrm{m^2/g}$) & $320-400$ & $198 - 258$ & $110 - 150$ \\
	Sodium ($\phi_m^{Na}$) & $0.006$ & $0.004$ & $0.002$ \\
	pH & $9.7 - 10.3$ & $9.2 - 9.9$ & $8.5 - 9.5$ \\
   	$R$ ($\mathrm{nm}$) & $3.3 - 4.1$ & $5.1 - 6.7$ & $8.8 - 12.1$ \\
 \hline
    \end{tabular}%
    \label{tab:ludox_spec}%
\end{table}%
Various techniques allow an accurate characterization of particle size and dispersity (see e.g. \cite{Orts-Gil2010,Teulon2018} for recent reviews). We resorted to SAXS technique. The SAXS signal depends on both the structure of the sample (\textit{i.e.}\ the correlations in particle positions) and its particle shape and size.  More formally, the scattered intensity  for an isotropic material $I(q)$ can be expressed as:
\begin{equation}
I(q) = S(q)P(q),
\label{eq:I_SP}
\end{equation}
with $S(q)$ the structure factor and $P(q)$ the form factor. $q$ is the norm of the scattering vector, and is expressed as
\begin{equation}
q = \frac{4\pi\un{sin(}\theta \un{)}}{\lambda}, 
\label{eq:q_expr}
\end{equation}
where $2 \theta$ is the scattering angle and $\lambda$ is the wavelength of the X-ray beam. 
In a dilute suspension, particles positions are uncorrelated and $S(q) = 1$. Thus, measuring the scattered intensity yields the form factor $P(q)$, from which particle size can be inferred.

The form factor for a monodisperse assembly of spheres with radius $R$ (as conjectured for our Ludox suspensions) can be written as:\cite{FeiSve87}

\begin{equation}
P_{M}(q|R) = 9 \frac{(\sin(q R)-qR \cos(qR))^2}{(q R)^6}
\label{eq:P_mono}
\end{equation}

\noindent Here, index $M$ stands for 'monodisperse' and the notation $f(x|a,b,...)$ denotes that $f$ is a function of  variable $x$, invoking $a$, $b$, etc as parameters. Equation (\ref{eq:P_mono}) is normalized so that $P_{M}(q=0|R) = 1$.

For a polydisperse suspension of spherical particles, $P(q)$ is the weighted sum of the individual form factors over the distribution of particle radii:\citep{Ped97}

\begin{equation}
P(q) = \displaystyle \int_{0}^{\infty} P_M(q|R)D(R) \, \mathrm{d}R,
\label{eq:P_poly}
\end{equation}

\noindent where $P_M(q|R)$ is given by eqn (\ref{eq:P_mono}) and $D(R)$ is the proportion of particles with radius between $R$ and $R+\mathrm{d}R$. Assuming that particle size follows a log-normal distribution, that is:

\begin{equation}
D(R|R_m,\sigma) = \frac{1}{R\sigma\sqrt{2\pi}} \: \exp\left(-\frac{(\un{ln}(R) - \un{ln}(R_m))^2}{2\sigma^2}\right),
\end{equation}

\noindent where $R_m$ is the median sphere radius and $\sigma$ is the standard deviation, the form factor then takes the form:

\begin{equation}
P(q|R_m,\sigma) = \displaystyle \int_{0}^{\infty} P_M(q|R)D(R|R_m,\sigma) \, \mathrm{d}R
\label{eq:P_fit}
\end{equation}

SAXS measurements were carried out at the LPS in Orsay on a homebuilt instrument operating with monochromatic copper radiation  (wavelength $\lambda = 0.154 \un{nm}$) delivered to the sample by a multilayer W/Si optics coupled with a rotating anode X-ray generator (Rigaku HU3R, $40 \un{kV}$-$40 \un{mA}$). Data were collected on a 2D pixel detector (Pilatus 250K, Dectris, Baden, Switzerland) with a pixel size of 0.172 mm using a sample-detector distance of $1208\pm2$ mm. The size of the X-ray beam on the sample was 0.8 x 0.8 $\un{mm^2}$.
 
\begin{figure}[ht!]
%\begin{figure}
\centering
\includegraphics[width=\columnwidth]{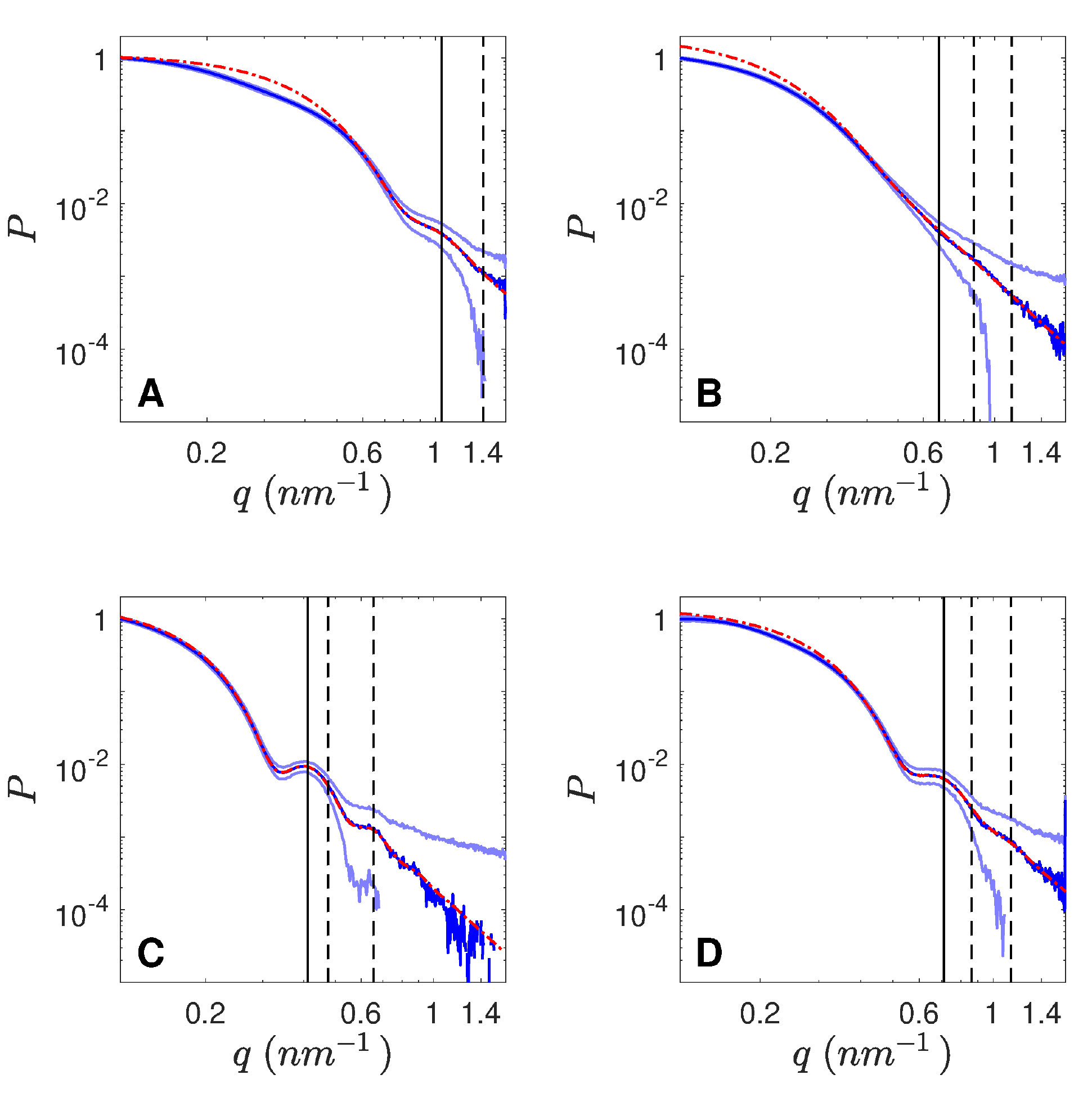}
\caption{Log-log plot of the measured form factor (average: in blue, standard deviation: in light blue) and fitted form factor (in red, dash-dot line) for the four Ludox suspensions used in this study: SM (A), HS-p (B), TM (C) and HS-m (D). In each panel, the full vertical lines indicate $q = 5.763/R_m$, where $R_m$ is the median particle radius obtained from the fits. This value for $q$ corresponds to a local maximum in $P(q)$ in the case of a monodisperse suspension. The two dashed vertical lines indicate $5.763/R^{spec}_{min}$ and $5.763/R^{spec}_{max}$ where $R^{spec}_{min}$ and $R^{spec}_{max}$ are the lower and upper bounds of the particle size range as inferred from the supplier specifications (line 5 in Tab.\ \ref{tab:ludox_spec}), except for panel (A) where the bar corresponding to $R^{spec}_{min} = 3.3\un{nm}$ is out of bounds at $q$ = 1.86 nm$^{-1}$.}
\label{fig:P_q_all}
\end{figure}

Dilute Ludox suspensions with silica volume fraction $\phi_v = 0.45 \%$ were prepared and NaCl was added to a $40 \un{mM/L}$  ($40 \un{mN}$) concentration in order to reduce electrostatic interactions between the particles. Glass capillary tubes (diameter 1 mm, wall thickness 0.01 mm) were filled with this dilute suspension and placed in the SAXS instrument for measurements (acquisition time $t_{acq} = 7200 \: \mathrm{s}$). A background measurement from a reference tube filled with 40 mN NaCl solution was obtained and subtracted from other measurements reported herein. Figure \ref{fig:P_q_all} displays the scattering profiles obtained for each of the four suspensions. In each of these graphs, a bump is visible; the position of this bump can be related to the inferred radii of the colloidal particles (row 5 in Tab.
\ref{tab:ludox_spec}). This bump is clear for Ludox SM, TM, and one of the two HS batches, but barely visible in the second HS batch. As will be seen later on, this illustrates the higher polydispersity of this latter suspension.  

Equation (\ref{eq:P_fit}) provides a quantitative estimate of the particle size dispersity. A significant source of resolution loss is the finite width of the X-ray beam, which introduces a smearing effect in the scattering pattern. The actual scattering pattern measured at the detector, $I(q)$, can be written as a convolution product: 

\begin{equation}
I(q) = I_{beam}(q) \circledast I_0(q), 
\end{equation}

\noindent where $I_{beam}(q)$ is the intensity profile of the incoming beam and $I_0(q)$ is the \textit{ideal} scattering pattern that would have been obtained using an incoming beam with zero width. The scattering pattern measured on a polycristalline calibration sample with sharp diffraction peaks then provides an estimate of the incoming beam width. Since the grain size in the polycrystalline sample is much larger than the wavelength of the X-rays, the scattering rings in $I_0(q)$ have negligible width, and the width of the rings in the measured pattern $I(q)$ equals the width of the beam. Scattered intensity as a function of radius $I(q)$ was computed from the scattering pattern measured on a home-made mesoporous SBA-15 silica powder\cite{LiForBle16} which is used as a calibration material at LPS. This material has three sharp Bragg diffraction peaks and a Gaussian fit on the first one yielded the full width at half maximum of the X-ray beam: $w_{beam} = 1.85 \un{mm}$. 

Intensity profiles $I(q)$ acquired on dilute suspensions were fitted by the convolution of $P(q|R_m,\sigma)$ as given by eqn (\ref{eq:P_fit}) with the inferred beam profile $I_{beam}(q)$.  This provides the median radius $R_m$ and standard deviation $\sigma$ of particle radius for the four suspensions used in this study. Table \ref{tab:ludox_sizing} summarizes these results. These values are fully consistent with results reported in literature. \cite{Orts-Gil2010,ZenGraOli11,Teulon2018} These literature values were obtained via SAXS measurements, \cite{Orts-Gil2010,ZenGraOli11,Teulon2018}, transmission electron microscopy (TEM), \cite{Orts-Gil2010,ZenGraOli11}, scanning electron microscopy (SEM)\cite{ZenGraOli11} and/or AFM. \cite{Orts-Gil2010,Teulon2018} 
Note that SAXS measurements provide better statistics than direct imaging using microscopy techniques since SAXS  performs measurements on a much larger number of objects: This number $N_{saxs}$ can be estimated from the silica volume fraction $\Phi_v=0.45\%$ of the analyzed suspensions, the sample volume $V_{tube}=1\un{mm}^3$ probed by the beam, and the inferred particle radius $R_m$: $N_{saxs}\approx 3\Phi_v V_{tube} / 4\pi R_m^3$.
This yields $N_{saxs}\sim 10^{12}$ particles, to be compared with the few hundreds of particles involved in TEM, SEM or AFM analysis.

The two HS-40 batches are found to exhibit significantly different form factors (Fig.\ \ref{fig:P_q_all}B and D) and thus different dispersities. The most monodisperse HS-40 sample ($\sigma/R_m = 0.14$), which was used in the previous studies,\cite{PirLazGau16} will henceforth be referred to as HS-m. The second and more polydisperse HS-40 sample ($\sigma / R_m=0.31$) is labelled HS-p henceforth. Comparing Tab.\ \ref{tab:ludox_sizing} column 3 with Tab.\ \ref{tab:ludox_spec}, row 5, the median radius $R_m$ calculated from experiments is always greater than the particle radius $R$ inferred from the specific area specified by the producer. This observation is consistent with literature.\cite{GauLazPau07,DiGDavMit12,PirLazGau16}.

\begin{table}[ht!]
%\begin{table}[htbp]
    \caption{Median particle radius and relative dispersity for the four Ludox suspensions used in this study. These values are extracted from SAXS diffraction patterns presented in Fig.\ \ref{fig:P_q_all} using eqn (\ref{eq:P_fit})}
  \centering
    \begin{tabular}{cccc}
    \hline
         Suspension & Batch number & $R_m \: \un{(nm)}$ & $ \sigma/R_m$ \\
          \hline 
    \SM & MKBP6397V & 5.5 & 0.19 \\   
    \HSm & BCBK7778V & 8.1 & 0.14\\     
    \HSp & STBF8427V & 8.6 & 0.31\\
    \TM & 05105EE & 14.0 & 0.10\\
    \hline
    \end{tabular}%
    \label{tab:ludox_sizing}%
\end{table}%

\subsection{Drying protocol}

\begin{figure}[ht!]
\centering
\includegraphics[width=\columnwidth]{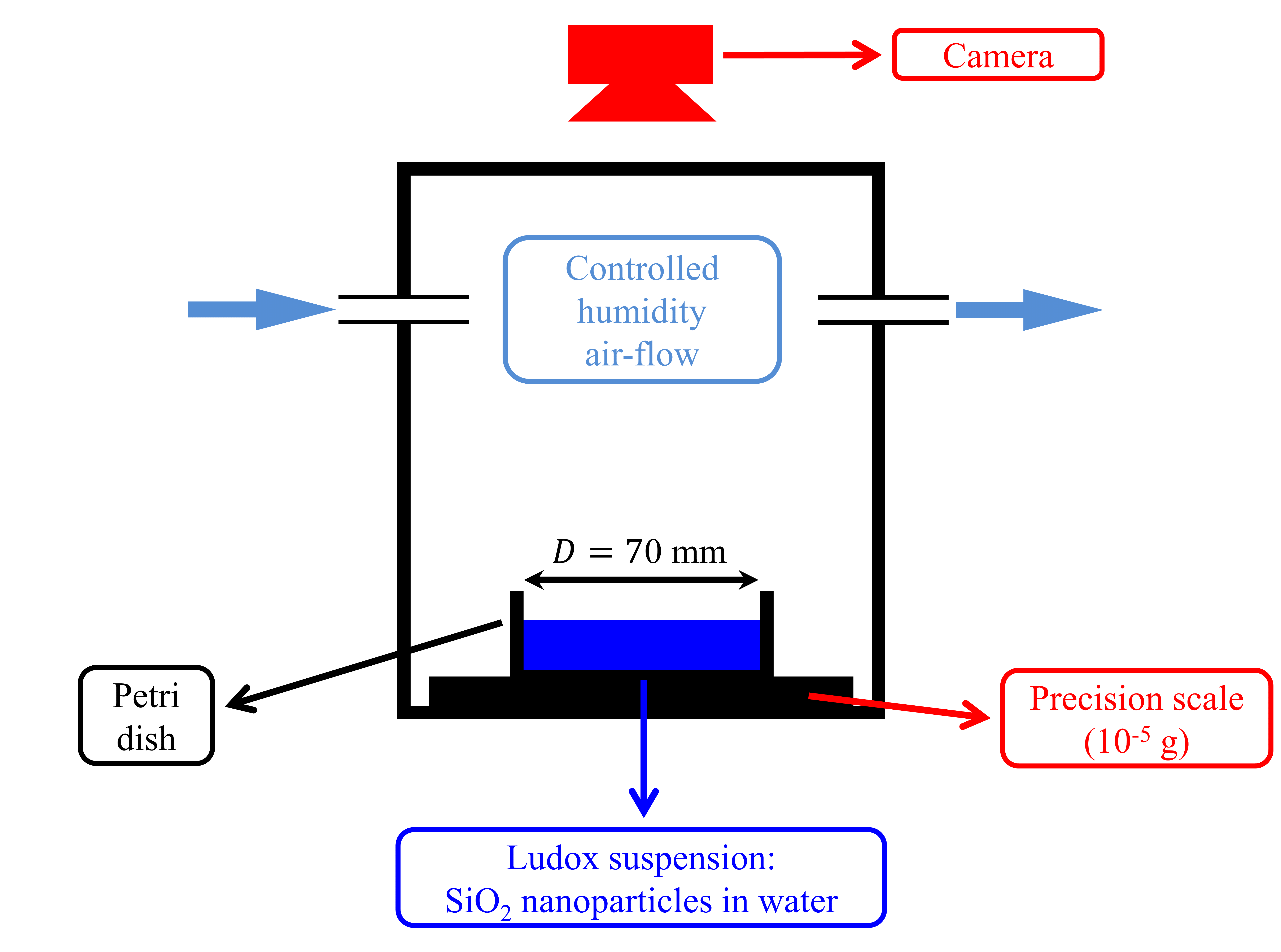}
\caption{Sketch of the drying setup used for the preparation of the colloidal layers. The colloidal suspension is poured in a Petri dish and left to dry under controlled humidity. A precision scale and a camera record the mass and visual evolution of the sample. The sketch is not at scale. Internal dimensions of the enclosure: $200\times 200 \times 260\,\mathrm{mm}^3$ (width $\times$ depth $\times$ height). The two pipes (diameter 1 mm) used to bring moist air in and out of the enclosure are placed $\sim 20\un{cm}$ above the Petri dish to limit the presence of a significant airflow sweeping the drying surface.}
\label{fig:drying_setup}
\end{figure}

Figure \ref{fig:drying_setup} presents a scheme of the drying setup. A mass $m_0$ of the considered suspension is first poured in a Glass Petri dish (diameter: $D_{petri} = 7 \un{cm}$); Petri dishes were beforehand washed with soap and rinsed with deionized water. The initial mass of suspension $m_0$ is chosen such that the final layer represents $10 \un{g}$ of dry solid material. Hence, $m_0 = 25 \un{g}$ for HS-m, \HSp and \TM suspensions (initial mass concentration $\phi_m = 40\%$) and $m_0 = 33.33 \un{g}$ for \SM suspensions (initial mass concentration $\phi_m = 30\%$).

The Petri dish rests on a precision scale  (Sartorius Cubis MSE225S, display resolution: $10^{-5} \un{g}$). A Labview routine, interfaced with the scale and a hygrometer, monitors the time evolution of the suspension mass, $m(t)$ and the temperature $T(t)$ and relative humidity $RH(t)$ in the enclosure. This routine additionally maintains $RH(t)$ around a prescribed value $RH_c$:  When $RH(t)$ drifts from $RH_c$ by more than $1\%$, a pump forces air from the scale enclosure either into a drying circuit (columns of silica gel, which absorbs moisture) or into a humidifying circuit (a water bottle in which the air stream bubbles).The presence of an airflow sweeping the drying surface could locally modify the vapor pressure and subsequently influence the drying behavior. To limit this effect, we used pipes of small diameter ($1\un{cm}$) to bring moist air in and out of the enclosure and placed them well above ($\sim 20\un{cm}$) the Petri dish. The entire setup (scale enclosure and pump circuits) is made airtight using glove box quality putty; sealing quality is especially important at low $RH$, as ambient air is a source of humidity. A camera placed above the enclosure and interfaced with a second Labview routine captures images of the drying colloidal layer at specified time intervals.

During the drying experiment, $m(t)$ decreases as water evaporates off the suspension. After some time, the mass of the drying layer stabilizes to its final value. Nevertheless, samples dried at high $RH_c$ may retain a significant amount of water in their pores (up to $25\%$ in mass). This may be due to capillary condensation at the nanopores in the solid layer. As water evaporates from the suspension, the free surface of the liquid gradually moves down; the corresponding speed, called the evaporation rate or drying rate, can be expressed as :

\begin{equation}
\dot{E}(t) = \frac{4\dot{m}(t)}{\rho_w \pi D_{petri}^2},
\end{equation}

\noindent where $\dot{m}(t) = \un{d}m(t)/\un{d}t$ is the mass loss rate and $\rho_w$ is the density of water. During the first phase of colloidal drying, the mass loss is nearly linear with time; that is, $\dot{m}$ and $\dot{E}$ are constants, respectively coined $\dot{m}_0$ and $\dot{E}_0$. In the rest of article, the expression "drying rate" will refer to the \textit{initial} drying rate $\Edot$.

\begin{figure}[ht!]
\centering
\includegraphics[width=\columnwidth]{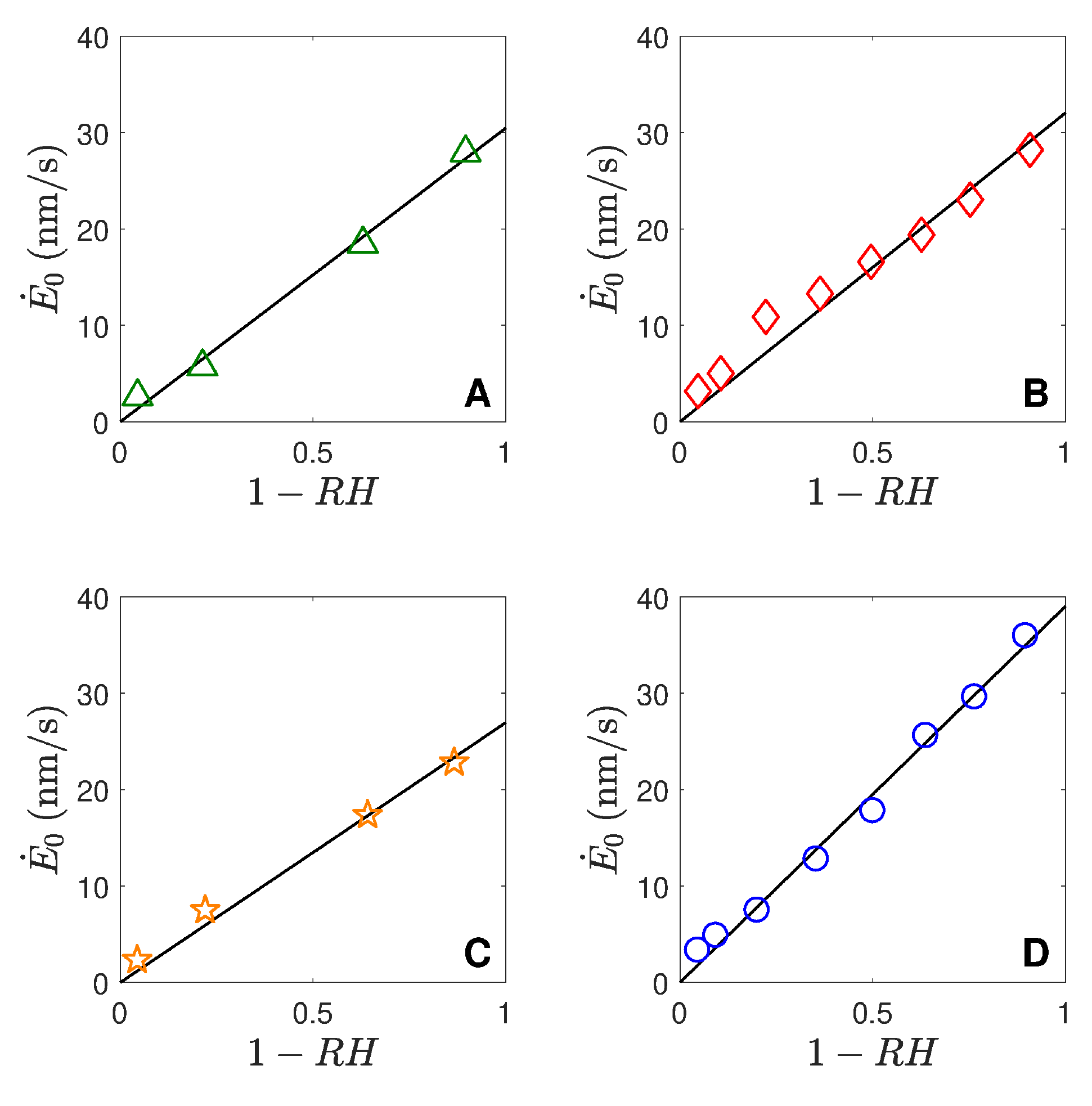}
\caption{Evaporation rate $\Edot$ as a function of relative humidity $RH$, for SM (A), HS-p (B), TM (C) and HS-m (D) layers. Black lines represent linear fits. The data for each suspension were fitted separately. Temperature variations from experiment to experiment explain the deviation of experimental data from the expected linear law.}
\label{fig:RH_Edot}
\end{figure}

In the dilute phase, the evaporation rate for the silica suspension is that of pure water. As such, it is expected to be proportional to $1-RH$.\citep{Cou00} This relation has been verified for the four suspensions, over the whole range of $RH$ explored (Fig.\ \ref{fig:RH_Edot}). Hereafter, $\Edot$ (rather than $RH$) is set as the control parameter. 
  
\subsection{Bulk packing fraction measurement}\label{sec:poro}

The packing fraction of the dry colloidal layers can be defined as the proportion of solid volume, $V_s$, to the total volume, $V_T$: $\phi = V_s/V_T$. 
The tightness of the particle packing can also be described using the porosity, $p$. This is defined as $p = 1 - \phi = V_p/V_T$, where $V_p$ is the pore volume (with $V_T = V_s + V_p$). Calculating $\phi$ requires to measure at least two parameters among the three ones: $V_s$, $V_p$ and $V_T$. In this study, $\phi$ was inferred from $V_s$ and $V_T$.

The volume $V_s$ occupied by the solid particles can be expressed as $V_s = m/\rho_s$, where $m$ is the mass of the dry sample and $\rho_s$ the density of the silica. The mass $m$ was measured after three hours of heat treatment at $200^\circ\mathrm{C}$, in order to evaporate off all water retained in the pores of the dried silica layer. In order to estimate the density of the silica particles, a dilution method, outlined in,\citep{CadHul02} was invoked on the \HSm suspension. Several dilutions of \HSm were prepared at prescribed mass fractions $\phi^{dil}_{m}$ and their densities $\rho_{dil}$ were measured. From the relationship $1-(\rho_{w}/\rho_{dil}) = \phi^{dil}_{m}(1-(\rho_{w}/\rho_s))$, where $\rho_w$ is the density of water, a linear regression yielded $\rho_s = 2.26 \pm 0.02 \un{g \: cm^{-3}}$. 

An accurate method for the determination of total volume $V_T$ invokes hydrostatic weighting.\cite{LesBonGau18} First, a sample of dried suspension was soaked in water.  At immersion, air bubbles are observed and a penetration front visibility progresses though the morsel. Samples were soaked for several hours prior to measurement, until these two processes ceased completly (meaning the morsel was visually homogeneous with stable mass). After soaking, the samples were  weighed in air (mass $m_{wet}$) and in water (apparent mass $\tilde{m}_{wet}$) using a standard hydrostatic weighing protocol. Prior to weighing, the water present on the sample surface was removed through careful wiping. As the water-soaked sample is left to dry in air, its measured mass decreases. This mass loss is quickly accompanied by a change in the visual appearance of the morsel, which become opaque. After soaking and blotting, by measuring the sample mass immediately, before opacification, we ensured the amount of water loss by evaporation was negligible. Tests showed the whole procedure to be well reproducible: Repeating the soaking and weighing procedure on few samples (mass $\sim 500-1000\un{mg}$), the same mass was measured within $1\un{mg}$, corresponding to less than $0.2\%$ mass variation.

Once $m_{wet}$ and $\tilde{m}_{wet}$ are measured, $V_T$ is deduced from Archimedes' principle: $V_T = (m_{wet} - \tilde{m}_{wet})/\rho_w$. Note that if the pores are only partially filled, this will equally affect $m_{wet}$ and $\tilde{m}_{wet}$ and cancel out in the expression of $V_T$. Thus, the accuracy of this method does not require the complete filling of the pores by the fluid, in contrast to other methods such as gas pycnometry or mercury porosimetry.

For each drying experiment, packing fraction measurements were repeated on two morsels. For \HSm layers, one morsel was soaked and weighted in water, and another one in ethanol. The density of the ethanol bath increased slightly with time; this is probably due to water absorption from the atmosphere. Porosity measurements on the layers obtained from the three other suspensions (\HSp, \SM and TM) were made using water for all the morsels.

\subsection{AFM imaging of the nanoscale particle arrangement at the layer surface}

In order to obtain spatial information on particle position, we used atomic force microscopy (AFM) to image the dried surfaces. A Bruker Dimension Icon Nanoscope V mounted with $5\un{nm}$ radius tips (MPP-11220-10, and then RTESPA-300 tips) provided images of the surfaces of the dried morsels in Tapping mode. For each drying experiment, a morsel was taken after drying and placed under the AFM head. To enable the averaging of surface properties, multiple images ($500\times 500 \un{nm}^2$, $512\times 512\un{pixels}^2$) were taken on the drying surface of each sample. As the samples are quite abrasive, tip wear occurs rapidly. The tip was thus replaced as soon as wear would degrade image quality, usually after 5 to 10 images.

\begin{figure}[ht!]
\centering
\includegraphics[width=.7\columnwidth]{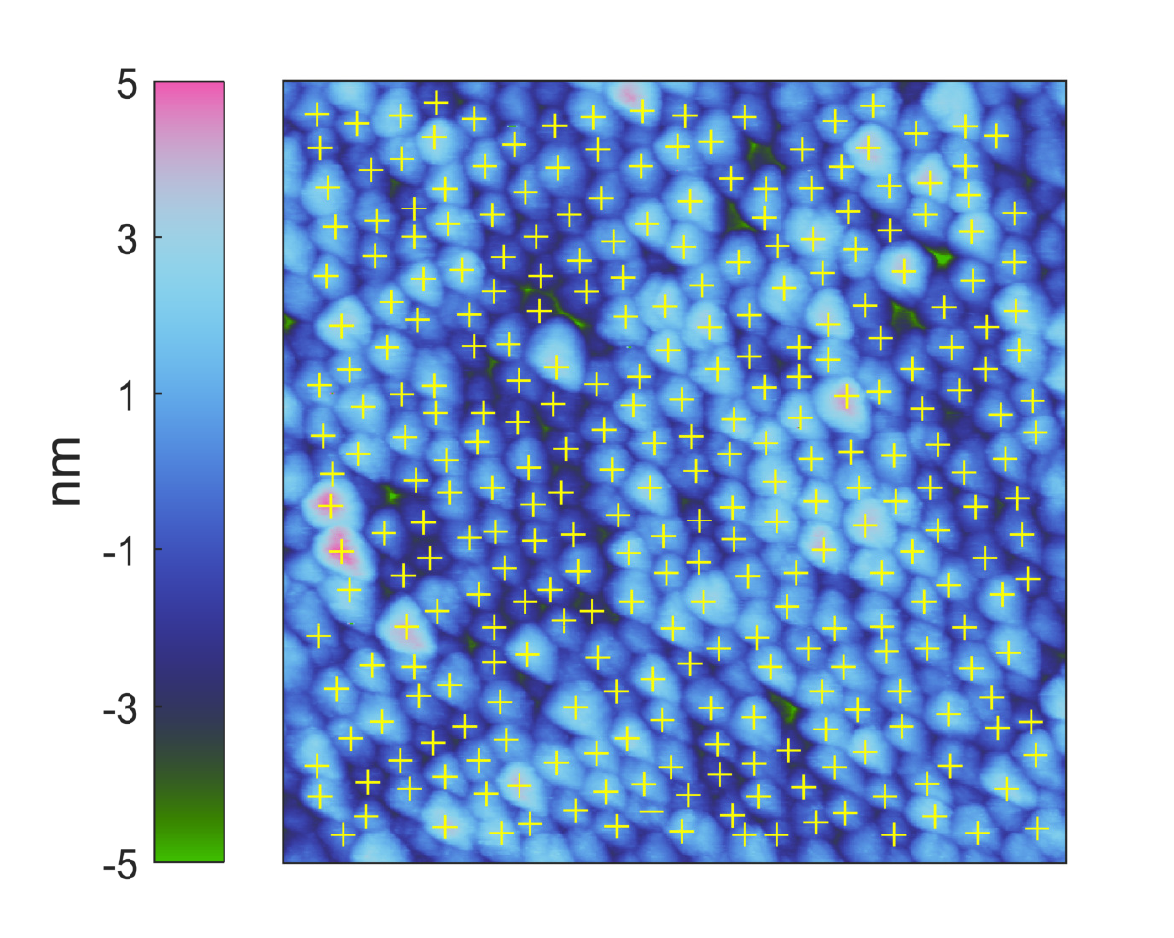}
\caption{Typical topographical AFM image, taken from the surface of a \TM layer dried at $RH = 95\%$. Image size is $500\times 500 \: \un{nm}^2$. The out-of-plane height ranges over 10 nm according to the colorbar on the right. Yellow crosses represent the positions of the particles centers as detected by the Matlab routine used in this study.}
\label{fig:afm_peaks}
\end{figure}

In order to compute statistics on the arrangement of the colloidal particles, it is necessary to extract the position of the particle centers from the topographical AFM images. To do so, a Matlab routine has been developed to allow the detection of all the local height maxima in a given image. Figure \ref{fig:afm_peaks} represents a typical output of this routine.

\section{Experimental results: Effect of drying rate and particle size distribution on dry layer nanostructure}

\subsection{Surface nanostructure: crystalline vs amorphous}

\begin{figure}
\centering
\vspace{.1cm}
\includegraphics[width=\columnwidth]{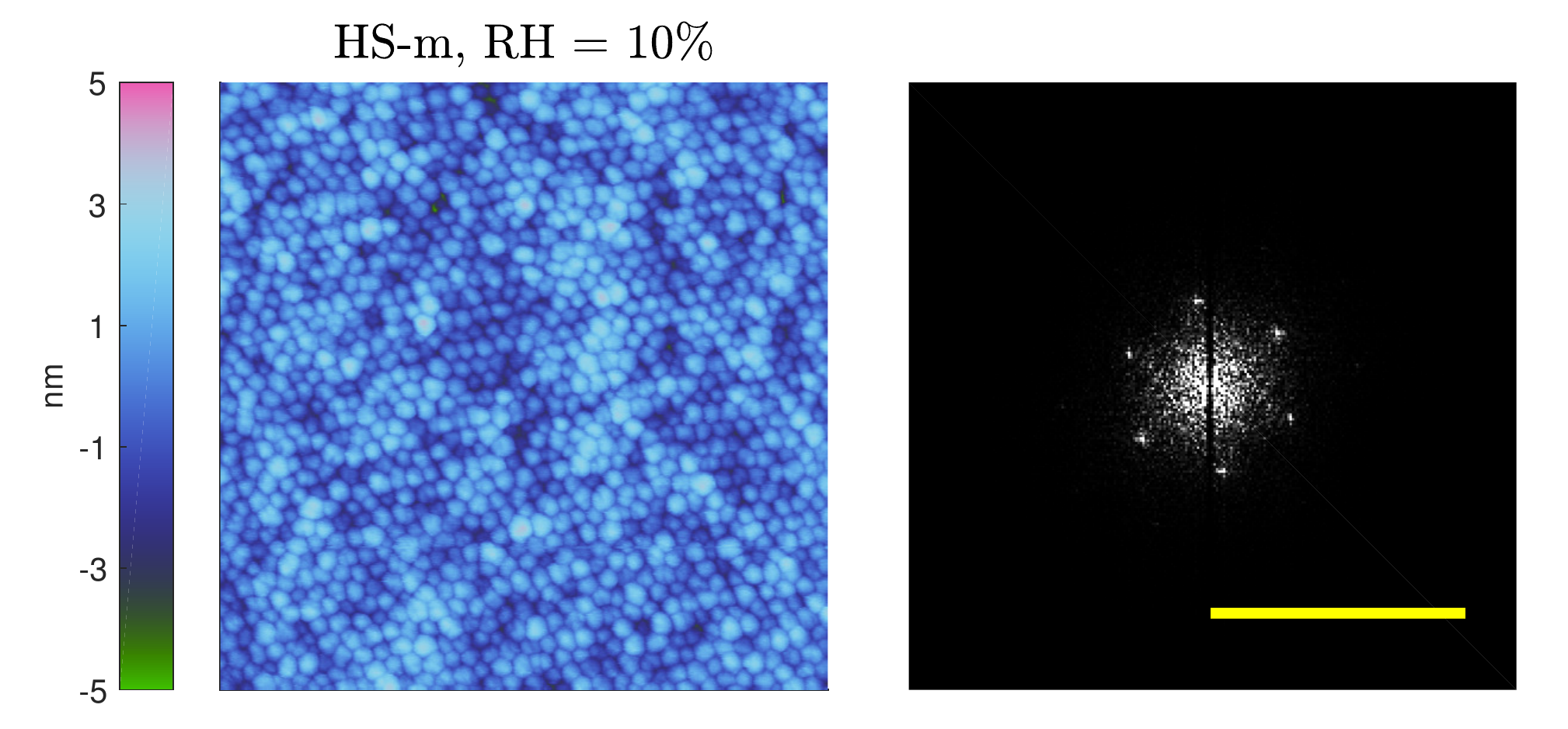}
%\vspace{.1cm}
\includegraphics[width=\columnwidth]{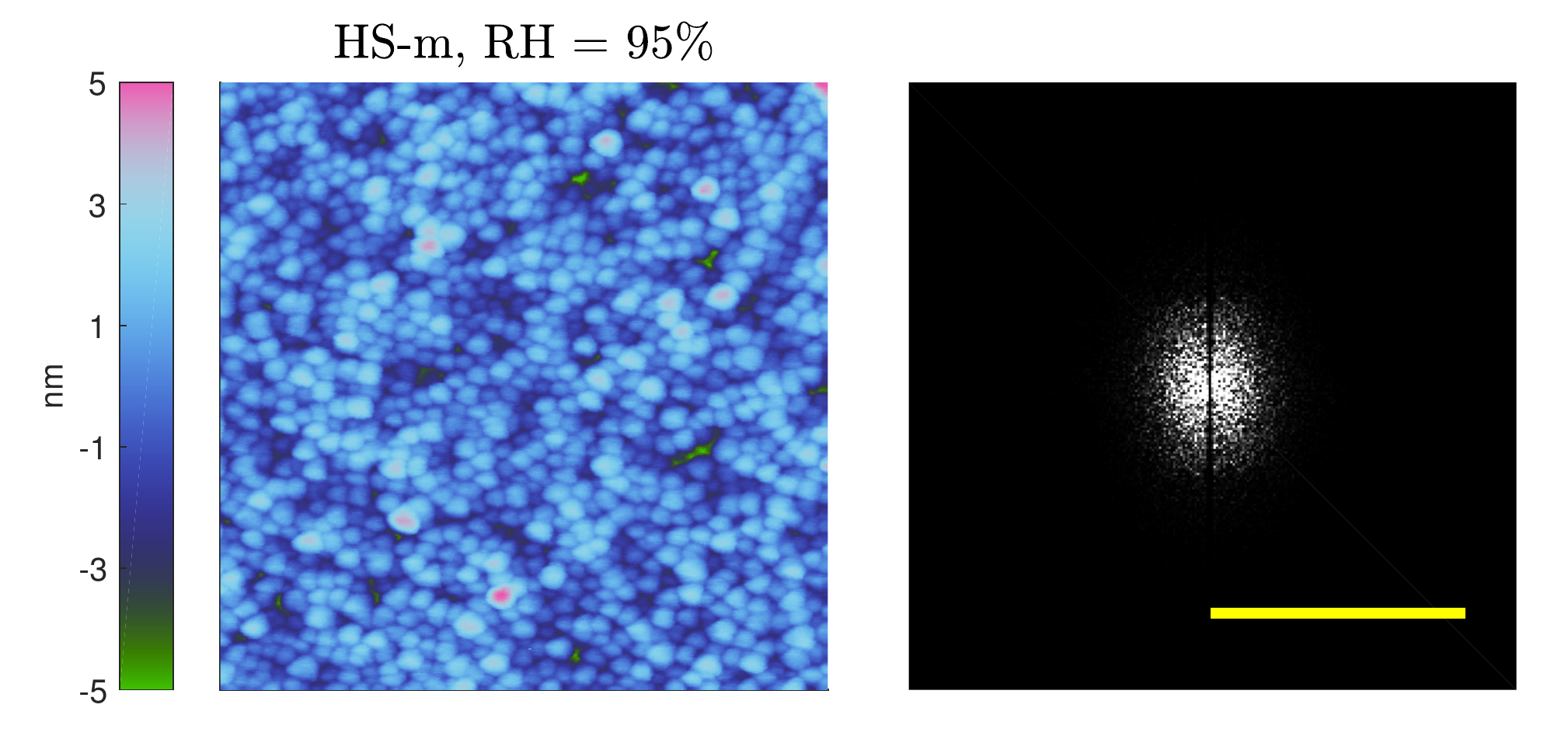}
%\vspace{.1cm}
\includegraphics[width=\columnwidth]{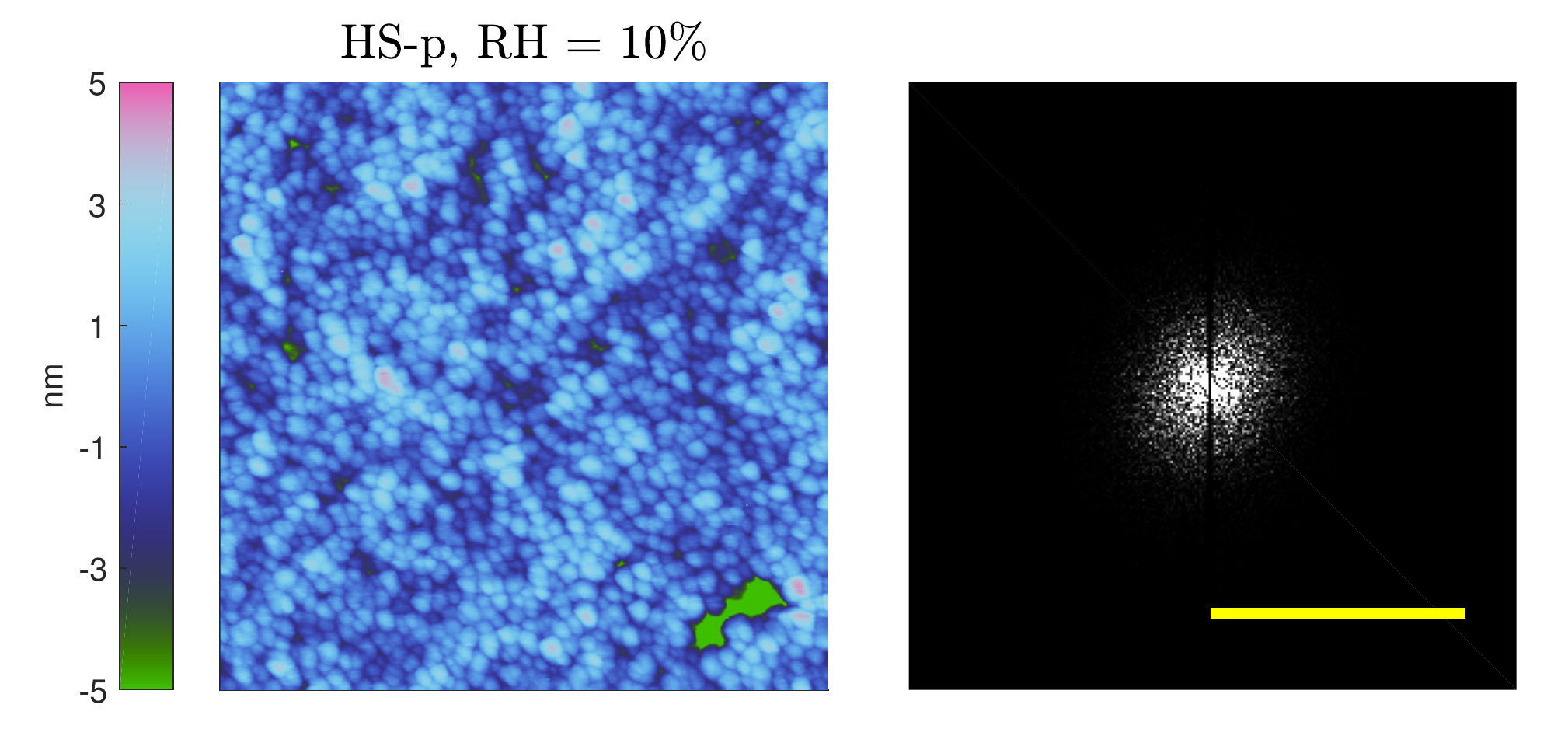}
%\vspace{.1cm}
\includegraphics[width=\columnwidth]{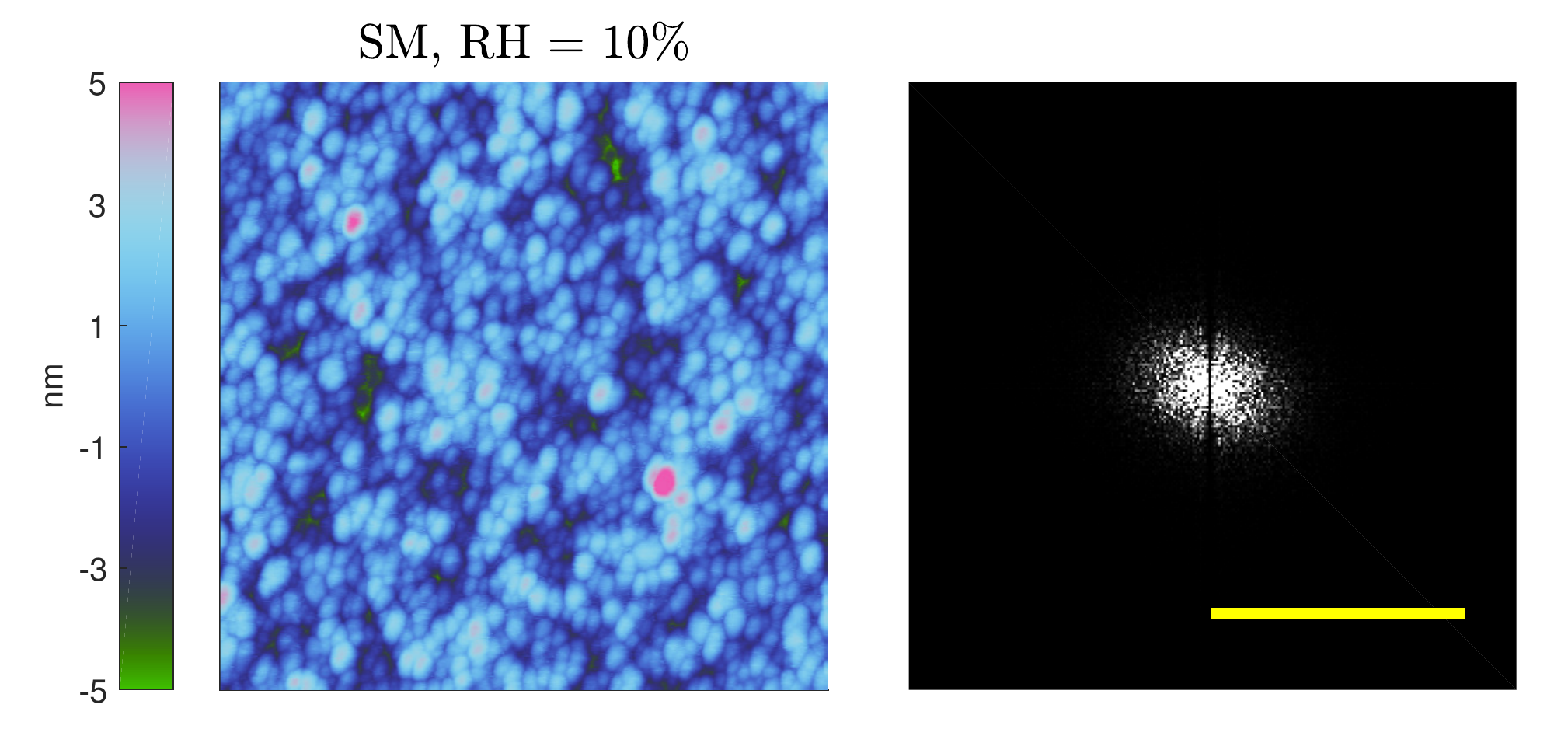}
%\vspace{.1cm}
\includegraphics[width=\columnwidth]{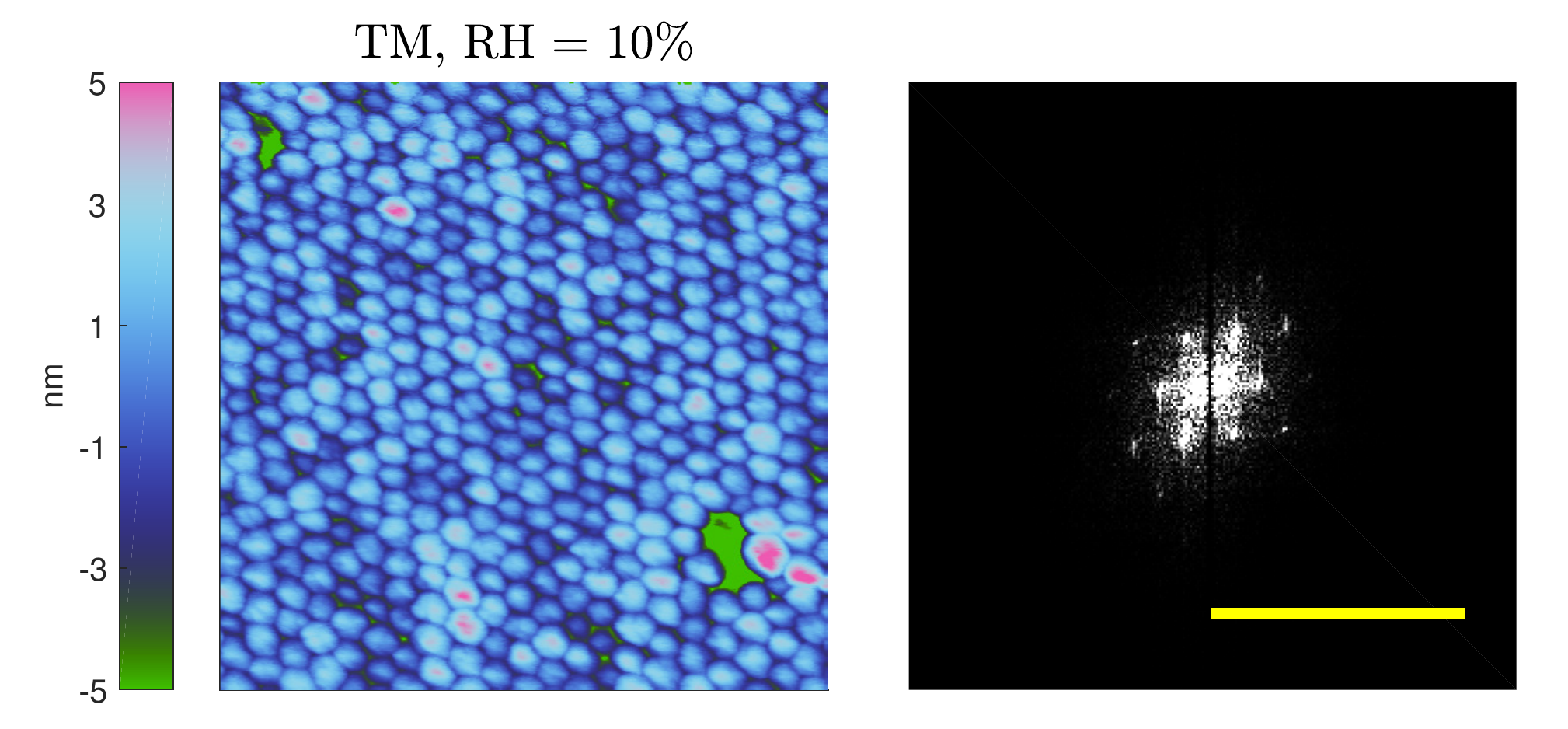}
\caption{$500$x$500 \: \un{nm}^2$ AFM images (left) and corresponding fast Fourier transforms (right) for representative layers. Yellow bar on FFT images represents $0.1 \un{nm^{-1}}$. Note that  AFM tip nominal radius ($5\un{nm}$) is of the same order as the colloid radius. Hence the topographies on the left-hand side result from the convolution of the real topography of the layer surface and the tip shape and cannot provide reliable quantitative measurements of colloid size.}
\label{fig:afm_all}
\end{figure}

Figure \ref{fig:afm_all}, left column presents typical AFM images of the dry silica layers surfaces. The corresponding 2-D Fourier spectra, shown on the right column, provide a direct visualization of the crystalline or amorphous nature of the arrangements. Depending on suspension type and drying rate, either a hexagonal structure or amorphous packings are observed. In short:
\begin{itemize}
\item Layers prepared from  \HSm suspension show a hexagonal surface pattern when dried at high evaporation rates (Fig.\ \ref{fig:afm_all}, row 1); layers dried at lower rates have an amorphous structure (Fig.\ \ref{fig:afm_all}, row 2).
\item Layers prepared from \SM (Fig.\ \ref{fig:afm_all}, row 3), as well as from \HSp suspension (Fig.\ \ref{fig:afm_all}, row 4) do not have visible crystallinity at any drying rate.
\item Layers prepared from  \TM suspension (Fig.\ \ref{fig:afm_all}, row 5) exhibit clear crystallinity at all drying rates. Note that both first-order and second-order diffraction peaks are visible on the 2D Fourier spectra.
\end{itemize}

\subsection{Surface nanostructure: translational \& rotational order}

Extracting the positions of the particle centers from the AFM images allows for a quantitative characterization of surface structure in the dry layers.

\begin{figure}[ht!]
%\begin{figure}[h!]
\centering
\includegraphics[width=\columnwidth]{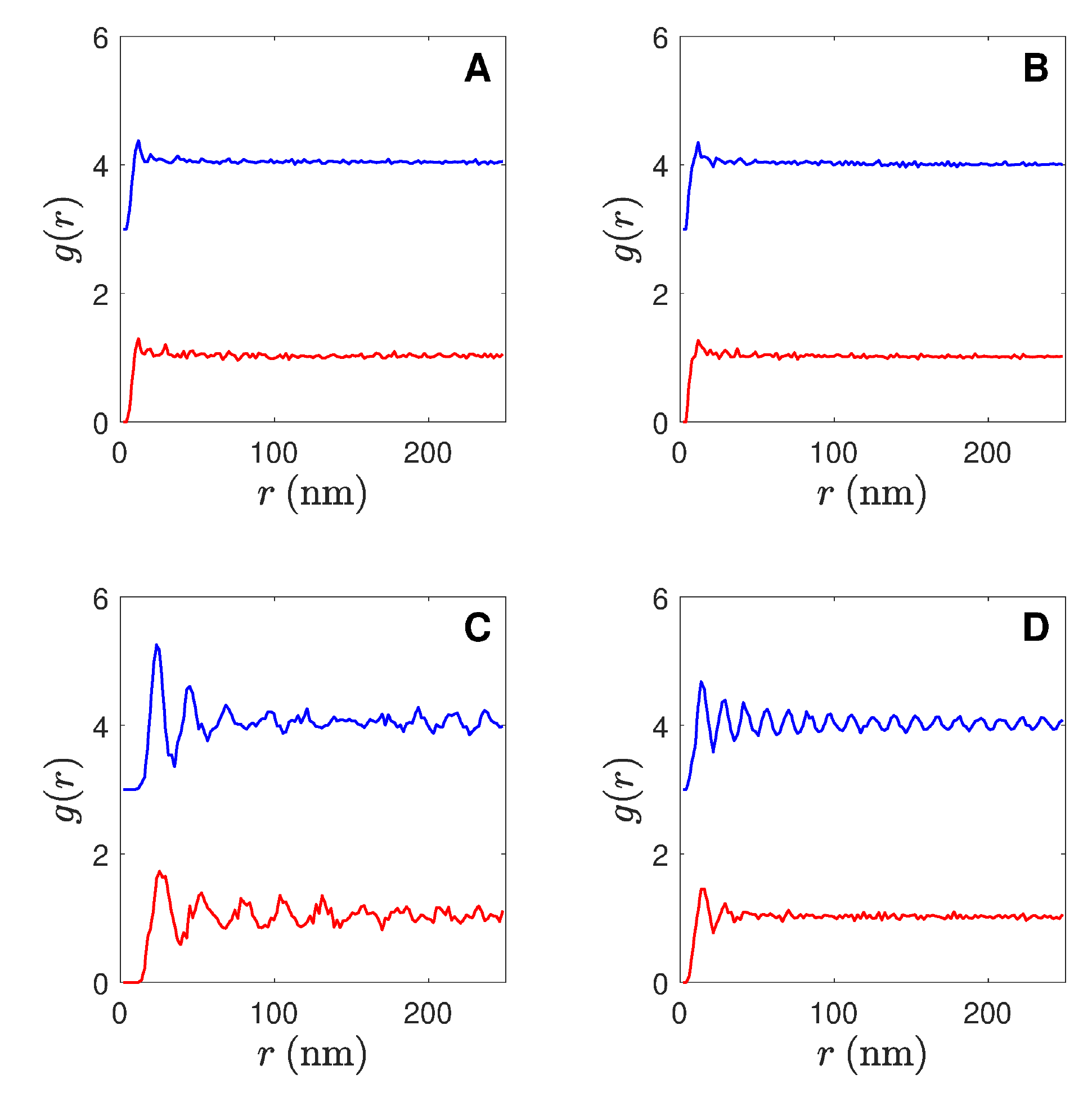}
\caption{Pair correlation function $g(r)$ at the highest drying rate (blue, top curves) and at the lowest drying rate (red, bottom curves) for SM (A), HS-p (B), TM (C) and HS-m (D) layers. For the sake of clarity, $g(r)$ was shifted upwards by three units for $RH = 10\%$. Note the high number of peaks observed on the curves in panel C and on the top curve of panel D. Those are indicative of the long-range translational order associated with a crystalline structure. They are absent in panels A and B, and in the bottom curve of panel D; the corresponding arrangements are amorphous.}
\label{fig:gr_all}
\end{figure}

Long-range translational order can be characterized using the pair correlation function $g(r)$. This function represents the probability of finding a particle center at a distance $r$ from a given particle, normalized by the probability for an ideal gas ($i.e.$ for random non-correlated particle positions).\cite{HansMcD13} Figure \ref{fig:gr_all} shows $g(r)$ for all four suspensions at maximal (blue, top curves) and minimal (red, bottom curves) drying rates. \TM layers and fast-dried ($RH = 10\%$) \HSm layers exhibit well-defined long-range peaks, which are the signature of long-range translational order. Conversely, $g(r)$ functions for slowly-dried ($RH = 95\%$) \HSm layers, as well as for \SM and \HSp layers, lack such peaks for $r$ larger than two particle diameters. Thus, no translational order exists at the surface of these layers.

\begin{figure}[ht!]
%\begin{figure}[H]
\centering
\includegraphics[width=\columnwidth]{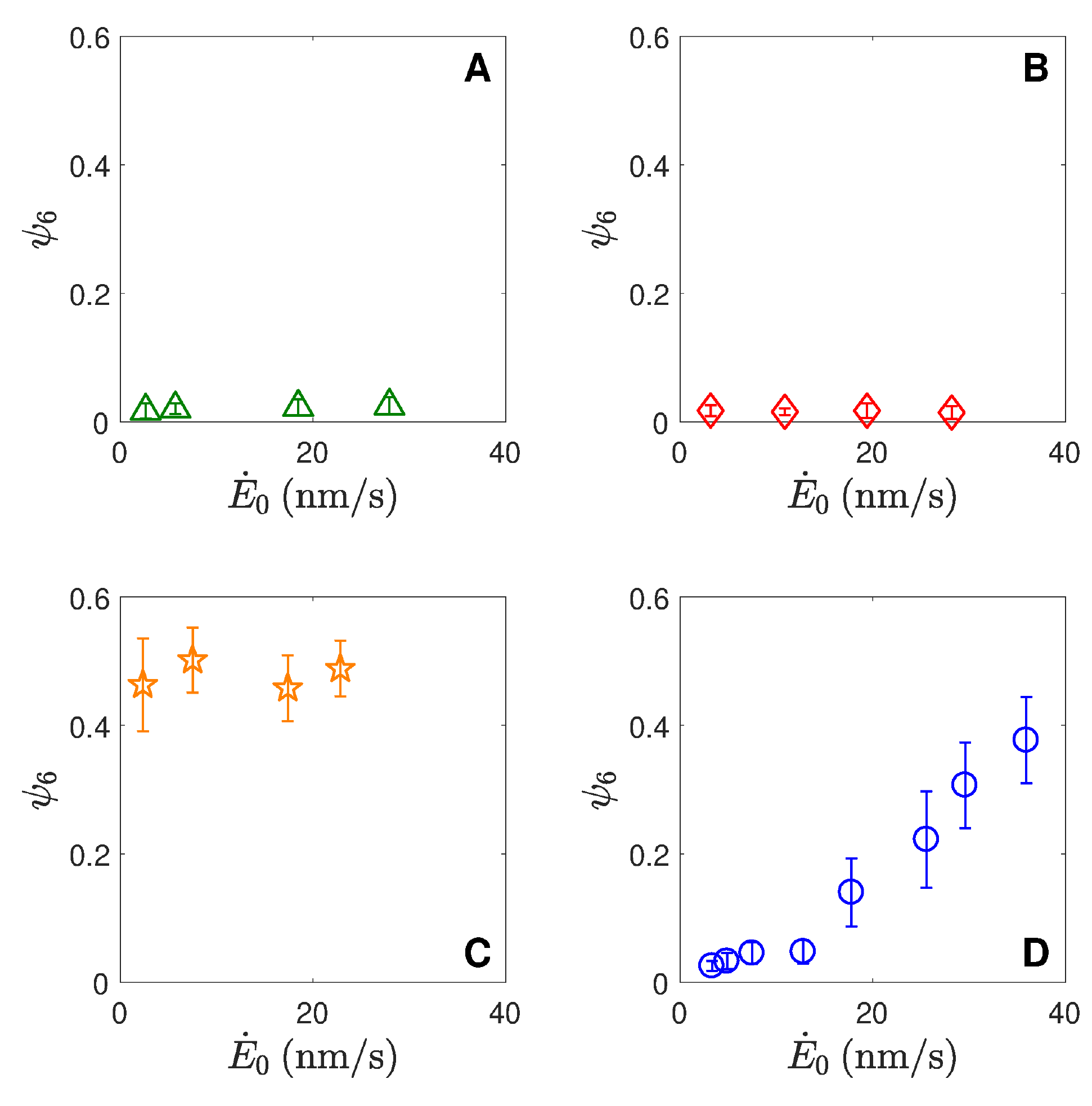}
\caption{Bond orientation parameter $\psi_6$ for the four Ludox types as a function of evaporation rate $\dot{E}_0$: SM (A), HS-p (B), TM (C) and HS-m (D) layers. The values remains always close to zero in panels A and B as expected for amorphous structures. They are finite and well above zero in panel C as expected for crystalline arrangements. Finally, a transition is observed in panel D; this is the signature of a rate-driven disorder-to-order transition.}
\label{fig:psi6_all}
\end{figure}

Crystallinity at the drying surface can be further characterized using the bond angle order parameter $\psi_n$, defined by:\cite{ReiIngSha06}

\begin{equation}
\psi_n = \left| \frac{1}{M}\sum_{k=1}^{M} \frac{1}{N_k}\sum_{l=1}^{N_k} \exp{(i \, n \, \theta_{kl})}  \right|,
\end{equation}

\noindent where $n$ is the number of nearest neighbors, $M$ is the number of particles in the AFM image, $N_k$ the number of nearest neighbors for particle $k$, and $\theta_{kl}$ is the angle between a fixed direction and the line joining particles $k$ and $l$. For a hexagonal arrangement, as observed on our layers, $n = N_k = 6$. $\psi_6$ is a measure of the orientational order of the particle arrangement. For an amorphous arrangement, $\psi_6 = 0$, while for an arrangement with sixfold symmetry (such as a perfect hexagonal lattice), $\psi_6 = 1$. Figure \ref{fig:psi6_all} gives $\psi_6$ as a function of the drying rate for each of the four Ludox suspensions. 

For all \SM and \HSp layers, $\psi_6 \simeq 0.05 \ll 1$ (Figs. \ref{fig:psi6_all}A and \ref{fig:psi6_all}B), which suggests that the layers are amorphous, irrespectively of the drying rate. This is in agreement with the $g(r)$ profiles on these layers, which show no crystalline peaks (Figs. \ref{fig:gr_all}A and \ref{fig:gr_all}B) beyond the next nearest neighbor. Conversely, $\psi_6 \simeq 0.47 \pm 0.03$ for \TM layers (Fig.\ \ref{fig:psi6_all}C), a value indicative of crystallinity. This is also consistent with the peaks observed in $g(r)$ for these layers (Fig.\ \ref{fig:gr_all}C). \HSm layers show two regimes, one with an amorphous structure ($\psi_6 \simeq 0.05$) at low evaporation rates ($\Edot \leq \dot{E}_c = 10 \un{nm/s}$), and one with increasing crystallinity for $\Edot > \dot{E}_c$, with $\psi_{6,max} = 0.38 \pm 0.07$ at the highest evaporation rate (Fig.\ \ref{fig:psi6_all}D). Even in the most crystalline layers, the particle centers are not perfectly aligned with the vertices of a hexagonal lattice. Moreover, thermal drift can introduce distortion in the AFM image, as can be seen for example on Fig.\  \ref{fig:afm_all}, row 5. Hence, $\psi_6$ is always significantly lower than one.

\subsection{Bulk packing fraction}

\begin{figure}[ht!]
%\begin{figure}[H]
\centering
\includegraphics[width=\columnwidth]{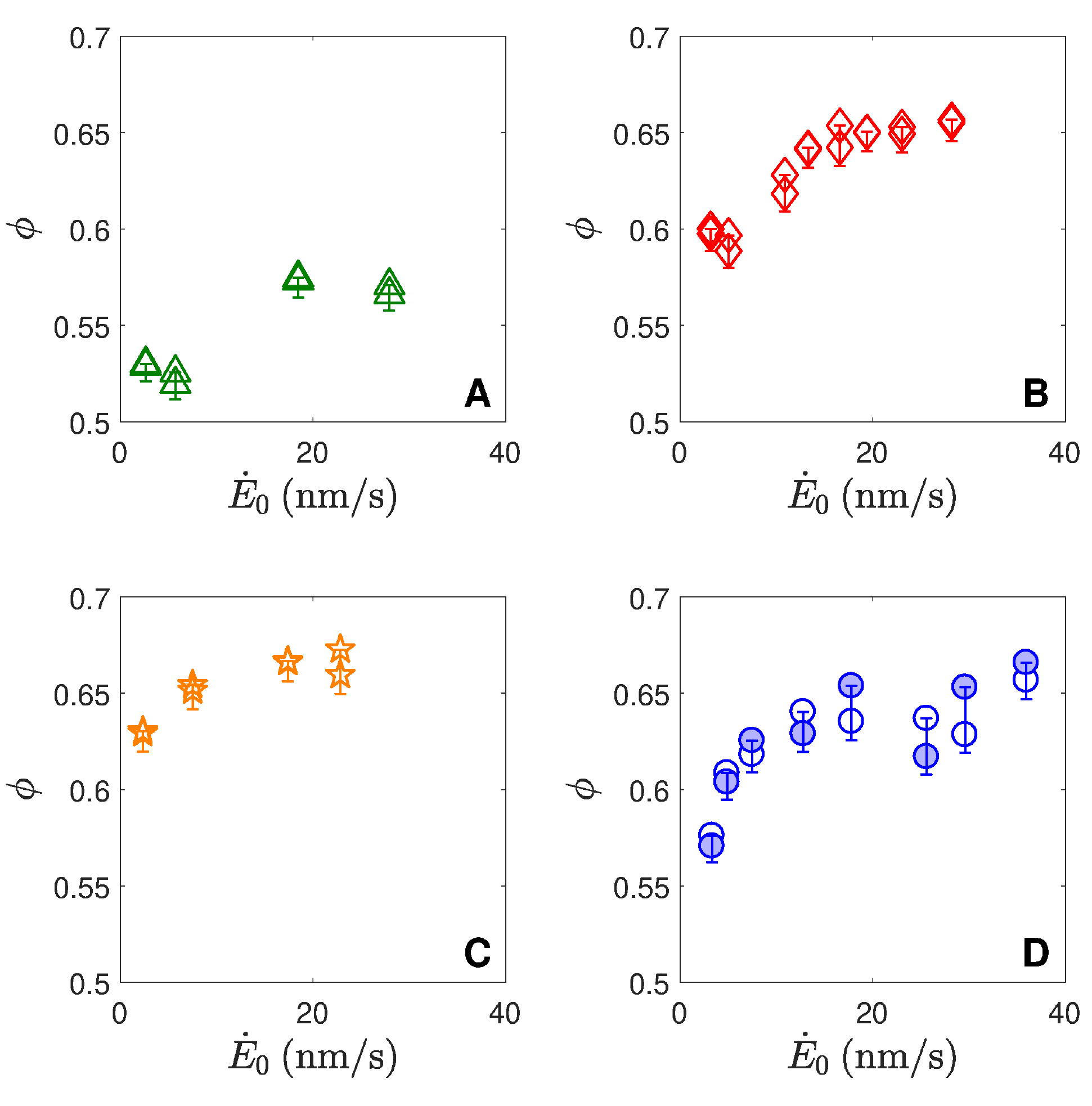}
\caption{Packing fraction ($\phi$) as a function of drying rate for all layers: SM (A), HS-p (B), TM (C) and HS-m (D). Measurements made on HS-m using ethanol for the immersion bath are shown with solid color. The axes limits are the same on each plot for easier comparison of the bulk packing fraction obtained when drying the different Ludox suspensions.}
\label{fig:phi_RH}
\end{figure}

The previous section concerned the characterization of  particle arrangement at the surface of the dried layer. We now turn to the analysis of  bulk structure via the measurement of packing fraction.  Figure \ref{fig:phi_RH} presents the packing fraction as a function of drying rate for the layers obtained from the four suspensions. Error bars are larger on the lower side, as they take into account the possible presence of water at the surface of the nanoparticles, which could overestimate $\phi$ by up to $1.5\%$.\citep{Ile79}

\begin{figure}[ht!]
%\begin{figure}[H]
\centering
\includegraphics[width=0.7\columnwidth]{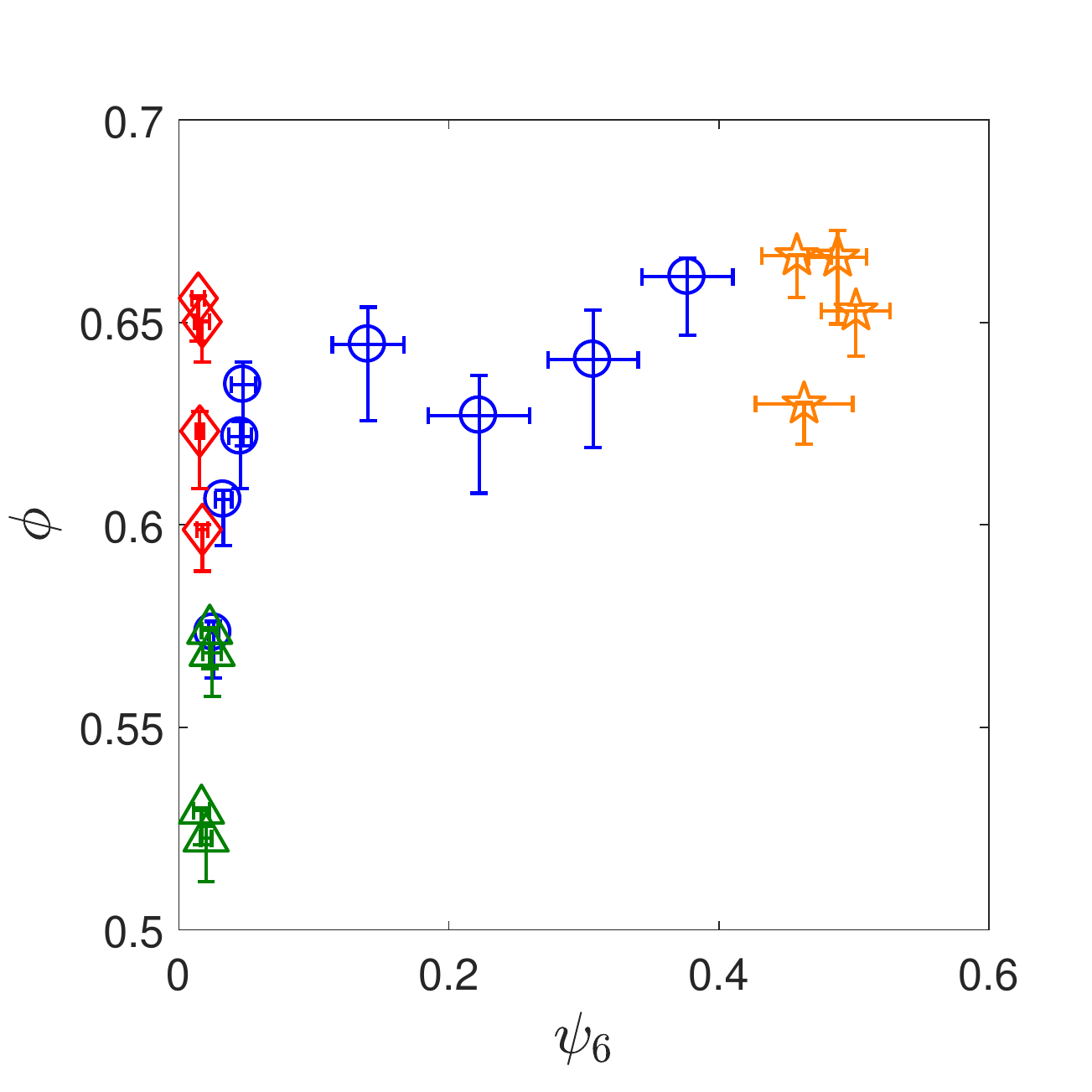}
\caption{Packing fraction $\phi$ as a function of bond orientation parameter $\psi_6$  for the four Ludox types : SM (green triangles), HS-p (red diamonds), TM (orange stars) and HS-m (blue circles).}
\label{fig:psi6_phi}
\end{figure}

In all cases, the packing fraction increases with increasing drying rate; hence, slowly dried layers are more porous. In \HSm and \HSp suspensions, a plateau is clearly observed  for $\Edot > 10 \un{nm/s}$. Volume fraction saturates at a value $\phi \simeq 0.65$, very close to the random close packing value ($\phi_{RCP}=0.64$ for monodisperse packing) and far below the value expected for 2D crystalline structures ($\phi_{HCC/FCC}=0.74$ for monodisperse packing). In other words, even in  \HSm layers which exhibits crystalline particle arrangement at the surface, the bulk is to a large extend amorphous, even at the highest drying rate. The range of volume fraction observed in \SM layers is much smaller, around $0.52 - 0.58$. Conversely, $\phi$ is much higher in \TM layers, up to $0.68$, which suggests that crystallization may extend significantly within the layer bulk. Figure \ref{fig:psi6_phi} gathers on a single graph the curves of $\phi$ as a function of $\psi_6$ for the four suspensions. It demonstrates the absence of univocal relationship between bulk volume fraction and surface arrangement: Crystalline surface arrangements are only observed in dense layers ($\phi \simeq 0.64-0.68$), while amorphous surface arrangements ($\psi_6 \simeq 0$) can be associated with either loose or dense packing, depending on evaporation rate and particle size and dispersity.  

In summary, this series of results show that:
\begin{itemize}
\item Low dispersity is required to observe surface crystallization: Crystallization is observed in all TM layers ($\sigma/R_m=0.1$) and on some HS-m layers ($\sigma/R_m=0.14$), but never on SM layers ($\sigma/R_m=0.19$) nor on HS-p layers  ($\sigma/R_m=0.31$). 
\item Crystallization is favored by increased evaporation rate, as observed on HS-m.
\end{itemize}
It is worth noting that this second observation may appear counter-intuitive; the slow evolution of a system generally favors order formation.  A possible explanation to this counter-intuitive result was advanced in,\cite{PirLazGau16} by postulating that slow evaporation favored the formation of aggregates, thus precluding particle crystallization.  The following section takes a closer look at this hypothesis by putting it to test.

\section{Evidence of particle aggregation at low evaporation rates}

\begin{figure}[ht!]
%\begin{figure}[H]
\centering
\includegraphics[width=\columnwidth]{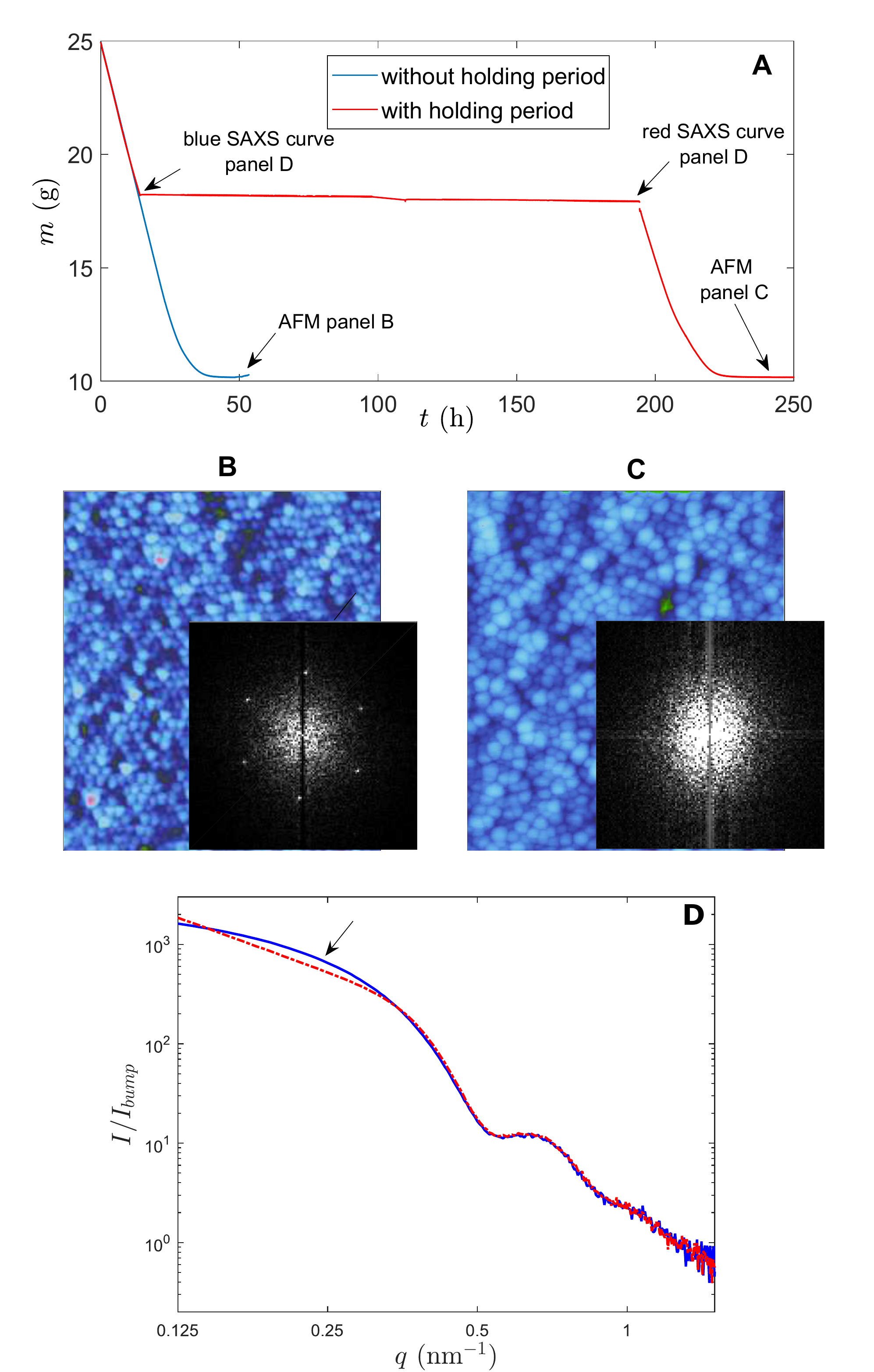}
\caption{
Influence of a holding period on the structure of the resulting dry layer. (A) Evolution of layer mass with time in a fast-drying experiment ($RH = 10 \%$) on \HSm suspension, with (red) and without (blue) a holding period. The holding period lasts $180 \un{h}$ at a packing fraction $\phi = 0.335$ (corresponding layer mass $m = 18.2 \un{g}$). (B-C) Fourier spectrum of a typical topographical AFM image of the dried layer obtained without holding period (left) and with a holding period (right) : The former presents a crystalline structure while the latter does not. (D) Log-log plot of the SAXS scattering intensity for the suspension (diluted at $\phi = 0.335$) at the beginning of the holding period (blue arrow in insert/blue curve) and at the end of the holding period (red arrow in insert/red curve).  The evolution of the shape of the signal at low $q$ (black arrow), \textit{i.e.}\ at large scales in direct space, demonstrates the formation of aggregates.
}
\label{fig:evol_plateau}
\end{figure}

Checking the postulate in Piroird et al.,\cite{PirLazGau16} we invoked SAXS measurements in conjunction with holding period experiments for the \HSm suspension, which was found to exhibit a disorder-to-order transition driven by the drying rate. During a holding period experiment, the suspension is first dried at the highest evaporation rate, in order to make the suspension more concentrated. The drying is then interrupted for a prescribed time, while the suspension is still in a liquid state  (Fig.\ \ref{fig:evol_plateau}A). Finally, the drying resumes at the original rate until completion. A fast-drying experiment yields a crystalline layer (Fig.\ \ref{fig:evol_plateau}B), yet a holding period experiment yields an amorphous one (Fig.\ \ref{fig:evol_plateau}C). Postulating the aggregate formation, this could be explained as follow: When the suspension remains long enough in a concentrated liquid state, whether during a slow-drying experiment or during a holding period, aggregates form. Once formed, they preclude the crystallization of the particles in the dried layer, regardless of the drying rate. 

SAXS measurements can now provide information on the actual formation of aggregates in the suspension during the holding period. The scattering curve of the particles in suspension was measured at the beginning and the end of the holding period. $25 \un{g}$ of \HSm suspension were dried at the highest evaporation rate ($\Edot = 24.2 \un{nm/s}$, $RH_c = 10\%$) down to a mass $m = 19 \un{g}$. $1.5 \un{g}$ of this concentrated suspension were sampled, diluted, and measured in SAXS, invoking the same dilution and measurement protocols as for the suspensions characterization (see Sec\ 2.1). The rest of the suspension was subjected to a holding period ($t = 168 \un{h}$); this aged suspension was then in turn diluted and measured in SAXS. 

Figure \ref{fig:evol_plateau}D represents the scattering curve of the sampled suspension, at the beginning of the holding period (solid blue line) and after the seven-day holding period (red dashed line). Before aging, $I(q)$ approaches a constant value at low $q$, which is indicative of a suspension of monodisperse particles. After the seven day holding period, $I(q)$ follows a power law at low $q$ (corresponding to a constant slope in log-log plot). A linear regression confirms that $I(q) \propto q^{-d}$ for q < 0.3 nm$^{-1}$, with an exponent $d = 1.85$. This is indicative of the formation of loose aggregates with a fractal dimension $d_f = 1.85$.\cite{MorParCab17} The evolution of the scattered intensity at low $q$ thus demonstrates the formation of aggregates in the concentrated suspension. This substantiates the scenario initially proposed in.\cite{PirLazGau16}

\section{Discussion}

This experimental study was designed to shed light on how colloidal particles self-assemble upon drying and form a solid layer of a given nanostructure. In this context, we examine bulk volume fraction and particle arrangement at the surface of solid layers obtained by drying millimeter-thick suspensions of silica nanospheres, using particle size and evaporation rate as control parameters. The principal results from our observations are: 

\begin{itemize}
\item[(i)] Either crystalline or amorphous particle arrangements can be observed at the surface of the dried layer: Amorphous structures are always observed in SM, crystalline structures are always observed in TM, and either crystalline or amorphous structures are observed in HS depending on drying rate and polydispersity. This observation can be compared with the reflectivity and scattering measurements in\cite{LeaLaiCou19} showing that the degree of ordering  increases with increasing particle size for the same types of Ludox suspensions. 
\item[(ii)] Crystallization requires both sufficiently high drying rate and low polydispersity. The two most polydisperse dispersions (\SM and \HSp) never crytallize. This is consistent with molecular dynamics simulations of hard spheres systems.\cite{PusZacVal09}. The threshold dispersity value for Ludox surface crystallization is observed to lie between 14 and 19 $\%$.
\item[(iii)] Crystalline surface arrangements imply dense bulk packing. Conversely, amorphous surfaces are observed in dry layers with either loose or dense bulk packing. 
\item[(iv)] When surface crystallization happen, the bulk volume fraction can reach values of up to $\phi \sim 0.68$, which is significantly larger than the random close packing value ($\Phi_{RCP}=0.64$ for monodisperse packings). Though polydispersity is known to increase $\Phi_{RCP}$, \cite{PhaRusZhu98,FarGro09} this increase is expected to be very small: $\Phi^{poly}_{RCP}(\sigma/R_m<0.15) \leq 0.655$ for a log-normal size distribution of dispersity $\sigma/R_m<0.15$.\cite{FarGro09} Hence, this alone cannot explain the value $\phi \sim 0.68$ observed in TM. This suggests that the crystallization is not limited to the surface, but extend through the bulk, even in these millimeter-thick layers. Note that, in the case of crystalline close packing,  polydispersity  make volume fraction smaller.\cite{PhaRusZhu98} This may explain why the observed value $\sim 0.68$ is smaller than the value $\Phi_{HCC/FCC} =0.74$ expected for crystalline closed packing of monodisperse spheres. 
\end{itemize}

\noindent Table \ref{tab:epl_summary} summarizes the interrelations between drying rate and particle size distribution on one hand, and surface order and bulk density on the other hand.

\begin{table}[ht!]
  \centering
    \caption{Summary of the influence of drying rate on surface structure and bulk packing fraction. $\phi$ values were sorted as low ($\phi<0.58$), medium ($0.58<\phi<0.63$) or high ($\phi>0.63$)}
{\footnotesize
    \begin{tabular}{ccccc}
	\hline
    Suspension & \SM & \HSm & \HSp & \TM \\
\hline
    Size, $R_m$ & $5.5\un{nm}$ & $8.1\un{nm}$ & $8.6\un{nm}$ & $14\un{nm}$ \\
    pH  & $9.7-10.3$ &9.2-9.9 & $9.2-9.9$ & $8.5-9.5$ \\
    Dispersity $\sigma/R_m$ & $0.19$ &0.14 & $0.31$ & $0.10$ \\
\hline
    Surface (low $\Edot$) & amorphous & amorphous & amorphous & ordered \\
    Surface (high $\Edot$)& amorphous & ordered & amorphous & ordered \\
\hline
   Bulk $\phi$ (low $\Edot$)& low & medium & medium & high \\
   Bulk $\phi$ (high $\Edot$) & low & high & high & high \\
\hline
    \end{tabular}
}%
  \label{tab:epl_summary}
\end{table} 

Peclet number, $Pe$, is usually considered as the relevant dimensionless parameter to rationalize how evaporation rate selects final packing structure. \cite{RouZim04,EkaMcdKed09} $Pe$ writes $Pe=h_0 \Edot/D_0$ where $h_0=5\un{mm}$ is the suspension thickness at drying initiation, and $D_0$ is the diffusion coefficient for the colloids. This number represents the ratio between Brownian diffusion time $\tau_{diff}=(h_0/D_0)^2$ and particle convection time toward the surface, which corresponds also to the layer evaporation timescale $\tau_{evap}=(h_0/D_0)^2$.
High $Pe$ indicates directional drying with a solid formed at the surface layer by layer, while low $Pe$ means isotropic bulk compression as water evaporates. To check whether or not $Pe$ controls final packing structure, we considered the experiments in \HSm.\cite{PirLazGau16} Using Stokes-Einstein relation, one gets $D_0\simeq 2.04 \times 10^{-11}\un{m}^2\mathrm{/s}$ in \HSm at $T=25\,^\circ\mathrm{C}$. From Fig. \ref{fig:psi6_all}D, experiments achieved with $Pe > Pe_c=1.5$ (i.e. $\Edot>\dot{E}_c=10\un{nm/s}$) should promote crystalline surface arrangement. This has been tested by performing an additional experiment \cite{PirLazGau16} in \HSm at a lower drying rate ($\Edot=26\un{nm/s}$), but starting with a suspension thickness twice larger ($h_0=10\un{mm}$). Peclet number in this additional experiment is $Pe=6.4$, i.e. much larger than $Pe_c$. Still, the particle arrangement was observed to be amorphous and not crystalline. This demonstrates that $Pe$ is not the relevant parameter driving particle packing in the experiments reported here. Beyond convection and diffusion captured by Peclet number, a third ingredient necessarly plays a key role in the selection of the final dried layer nanostructure.

The holding period experiment and SAXS measurements reported in Sec. 4 demonstrate that slow drying allows the formation of particle aggregates in the liquid phase. We argue this flocculation process to be this missing ingredient. Crystallization will be observed provided:
\begin{itemize}
\item[(i)] low enough polydispersity;
 \item[(ii)] flocculation is precluded before solidification.
 \end{itemize}
 Requirement (i) is fulfilled in \TM and \HSm suspensions since crystallization is observed for some drying conditions. Requirement (ii) is fulfilled in these two suspensions for fast enough drying, when evaporation timescale, $\tau_{evap}$,  is smaller than aggregation timescale, $\tau_{aggr}$. In \HSm, this is ensured for $\Edot>\dot{E}_c=10\un{nm/s}$, i.e. for $\tau_{evap} < 140\un{h}$. This value provides an order of magnitude for aggregation timescale in \HSm: $\tau_{aggr}(\mathrm{\HSm}) \approx 140\un{h}$. This is consistent with the observation that introducing an holding period of $180\un{h} > \tau_{aggr}(\mathrm{\HSm})$ 
 in a fast drying experiment promotes the aggregate formation, and subsequently precludes packing order at the dried layer surface.  In \TM, crystallinization is observed over the whole range of explored $\Edot$, down to $\dot{E}_{min} = 2.3\un{nm/s}$. This means that aggregation time in \TM is larger than $\dot{E}_{min}/h_0$: $\tau_{aggr}(\mathrm{\TM}) > 600\un{h}$. The fact that $\tau_{aggr}(\mathrm{\TM}) > \tau_{aggr}(\mathrm{\HSm})$ can be understood since colloids in \TM are larger. Hence, the diffusion coefficient $D_0$ is smaller, diffusion time $\tau_{diff}$ is larger, and finally, as $\tau_{aggr}$ increases with $\tau_{diff}$,\cite{LinLinWei89} aggregation time is larger. In \SM, no crystallinisation is observed over the whole range of explored $\Edot$, up to $\dot{E}_{max}=27.9\un{nm/s}$. This means that either requirement (i)  or (ii) are not fulfilled (or both): Either the polydispersity $\sigma/R_m=0.19$ is too high to allow crystallization, or the aggregation time is too small, $\tau_{aggr}(\mathrm{\SM}) < 50\un{h}$, to allow preclude flocculation in our drying experiments. Finally, in HS-p, no crystallinization is observed. The main difference between HS-p and HS-m is that polydispersity is much larger in HS-p ($\sigma/ R_m=0.31$, compared to $\sigma/ R_m=0.1$ for \HSm); hence the absence of crystallinisation in HS-p is due to the failure of requirement (i).

Variations in volume fraction provide another signature of flocculation. Large aggregates cannot pack very densely and must yield more porous layers. Consequently, large volume fractions are expected in  \TM since flocculation is precluded over the whole range of explored $\Edot$. Conversely, the small volume fractions observed in \SM at all $\Edot$ suggest that aggregation is important over the whole range of explored drying rate; this suggests that  the failure of requirement (ii) rather than failure of requirement (i) is responsible for the absence of crystallinization in \SM. For intermediate cases (\HSm and \HSp), layer density depends on the aggregate formation and is mediated by the drying rate. Note that in Ludox HS, dispersity does not have a significant effect on packing fraction: the porosities observed in \HSm and \HSp are roughly the same. It may be important at larger particle sizes, for which flocculation seems to be prevented. In the \TM suspension, indeed, volume fraction is  significantly larger than the RCP value ($\phi \simeq 0.68$ to be compared with $\phi_{RCP}=0.64$ in monodisperse packings); this suggests that surface ordering extends partially through the layer bulk. The way order and disorder coexist in the 3D bulk remains an open question. In this context, it is worth noting that a similar phenomenon of packing ordering promoted by faster colloid velocity has been reported by Noirjean et al.\cite{NoiMarDev17} during unidimensional drying in a 2D Hele-Shaw cell opened on one side. Direct imaging of the colloid arrangement during drying was possible in such a 2D geometry, and it was evidenced a succession of ordered and disordered domains with thickness of $\sim 100$ particle size. Similar organization may occurs in our 3D drying geometry. TEM imaging of the layer cross-section could provide further information on this aspect.\cite{YotYoo93}

\section{Conclusion}

The series of experiments reported here investigated how particles self-assemble during colloidal drying to form thick solid layers with varying nanostructures. In dispersions with hard, non deformable particles, such as the silica suspensions examined here, particle arrangement in the dried layer is demonstrated to be mediated by flocculation. This process occurs in the liquid phase. Aggregate formation depends on the drying rate, as reported in:\cite{PirLazGau16} slower evaporation leaves more time for flocculation to proceed and thus allows larger aggregates to form. 
In the systems and for the drying conditions examined here, final packing structure is mainly  driven  by the ratio of aggregation time over evaporation/convection time, rather than by the Peclet number (Brownian diffusion time over convection time) as is sometimes the case. \cite{RouZim04,EkaMcdKed09}
 This highlights that in general, both ratios shall be considered when the packing structure is investigated. Selection of aggregation time in presence of evaporation remains an open question: It is expected to increase with diffusion time\cite{LinLinWei89} and, as such, to increase with colloid size. It should also depend on the dispersion stability at  the onset of evaporation. Future measurements of the zeta potential during drying may shed a new light on this aspect. 
 Indeed, starting with Smoluchowski coagulation equation \cite{Smo17} and Derjaguin-Landau-Verwey-Overbeek (DLVO) theory, \cite{DerLan41, VerOve47} a series of increasingly sophisticated models \cite{Spie70,McgPar67,HonRoeWie71,BreWalVli95,GraSubBut02,LacSauRai19} has been developped to assess colloidal stability and flocculation kinetics (see e.g. \cite{RusSavSch12} for recent reviews). They may allow an estimation of the aggregation time from the knowledge of the ion concentration, zeta potential and Hamaker constant.
 
Depending on the extent of flocculation, dried layers of variable density are obtained : the smaller the aggregates, the higher the bulk volume fraction. When flocculation is precluded and the initial particle dispersity is low enough (less than $15\%$), crystallization is observed at the layer surface. As demonstrated here in Ludox TM, crystalline arrangement of the particle can also extend in 3D through the layer bulk, even in millimeter-thick dry layers.  

These results may be of interest in the design of processes requiring the auto-assembly of colloidal particles into large-scale crystalline structures, for example for photonics and biotechnology applications. They also shed a new light on why increasing drying rate yields more cracks in the dried layer:\cite{GauLazPau10,BouGioPau14} because of flocculation, slower drying yields more porous layers. Moreover, the elastic moduli of these layers is a decreasing function of volume fraction.\cite{LesBonGau18} Then, slower drying yields softer layers, which can accommodate capillary and/or substrate induced tension more easily, without cracking. This scenario is of course to be confirmed, as the crack patterns depends on the balance between available elastic energy and fracture costs\cite{GauLazPau10, BouGauLaz13, Laz17} in which many other mechanisms can be involved (chemical bond tightening increasing with time or dependence of the stress evolution with  drying rate\cite{CheLaz13} for instance). Work in this direction is under progress.

\section*{Conflicts of interest}
There are no conflicts to declare.

\section*{Acknowledgements}
The co-authors of this work would like to acknowledge financial support from Triangle de la Physique (RTRA), Ile-de-France (C'Nano, ISC- PIF, and DIM-Map), LabEx LaSIPS (ANR-10-LABX- 0040-LaSIPS) and PALM (ANR-10-LABX-0039-PALM) managed by the French National Research Agency under the "Investissements d'avenir" program (ANR-11-IDEX-0003-02).

%%%END OF MAIN TEXT%%%


%%%REFERENCES%%%
\providecommand*{\mcitethebibliography}{\thebibliography}
\csname @ifundefined\endcsname{endmcitethebibliography}
{\let\endmcitethebibliography\endthebibliography}{}

\end{document}